\documentclass[twocolumn,showpacs,preprintnumbers,amsmath,amssymb,prb,superscriptaddress]{revtex4}

\usepackage{epsfig}
\usepackage{graphicx}
\usepackage{dcolumn}
\usepackage{bm}

\begin{document}

\title{Field-- and pressure--induced magnetic quantum phase transitions 
in TlCuCl$_3$}

\author{Masashige Matsumoto} 
\affiliation{Theoretische Physik, ETH--H\"onggerberg, CH--8093 Z\"urich,
Switzerland}
\affiliation{Department of Physics, Faculty of Science, Shizuoka University,
836 Oya, Shizuoka 422--8529, Japan}

\author{B. Normand}
\affiliation{D\'epartement de Physique, Universit\'e de Fribourg, CH--1700 
Fribourg, Switzerland}

\author{T. M. Rice}
\affiliation{Theoretische Physik, ETH--H\"onggerberg, CH--8093 Z\"urich, 
Switzerland}

\author{Manfred Sigrist}
\affiliation{Theoretische Physik, ETH--H\"onggerberg, CH--8093 Z\"urich, 
Switzerland}

\date{\today}

\newcommand{\br}{{\mbox{\boldmath$r$}}}
\newcommand{\bh}{{\mbox{\boldmath$h$}}}
\newcommand{\bH}{{\mbox{\boldmath$H$}}}
\newcommand{\bk}{{\mbox{\boldmath$k$}}}
\newcommand{\sk}{{\mbox{\footnotesize $k$}}}
\newcommand{\bsk}{{\mbox{\footnotesize \boldmath$k$}}}
\newcommand{\bsq}{{\mbox{\footnotesize \boldmath$q$}}}
\newcommand{\bsQ}{{\mbox{\footnotesize \boldmath$Q$}}}
\newcommand{\bsr}{{\mbox{\footnotesize \boldmath$r$}}}
\newcommand{\bS}{{\mbox{\boldmath$S$}}}
\newcommand{\bQ}{{\mbox{\boldmath$Q$}}}
\newcommand{\bd}{{\mbox{\boldmath$d$}}}
\newcommand{\bsd}{{\mbox{\footnotesize{\boldmath$d$}}}}
\newcommand{\bsigma}{{\mbox{\boldmath$\sigma$}}}
\newcommand{\Tl}{{TlCuCl$_3$}} 
\newcommand{\K}{{KCuCl$_3$}} 
\newcommand{\N}{{NH$_4$CuCl$_3$}} 
\newcommand{\ha}{{\hat{a}}} 
\newcommand{\hb}{{\hat{b}}} 
\newcommand{\hc}{{\hat{c}}} 
\newcommand{\hx}{{\hat{x}}} 
\newcommand{\hy}{{\hat{y}}} 
\newcommand{\hz}{{\hat{z}}} 
\newcommand{\muB}{{\mu_{\rm B}}} 
 
\begin{abstract} 
 
Thallium copper chloride is a quantum spin liquid of $S$ = 1/2 Cu$^{2+}$ 
dimers. Interdimer superexchange interactions give a three--dimensional
magnon dispersion and a spin gap significantly smaller than the dimer
coupling. This gap is closed by an applied hydrostatic pressure of 
approximately 2kbar or by a magnetic field of 5.6T, offering a unique 
opportunity to explore the both types of quantum phase transition and 
their associated critical phenomena. We use a bond--operator formulation 
to obtain a continuous description of all disordered and ordered phases, 
and thus of the transitions separating these. Both pressure-- and 
field--induced transitions may be considered as the Bose--Einstein 
condensation of triplet magnon excitations, and the respective phases 
of staggered magnetic order as linear combinations of dimer singlet 
and triplet modes. We focus on the evolution with applied
pressure and field of the magnetic excitations in each phase, and in
particular on the gapless (Goldstone) modes in the ordered regimes which
correspond to phase fluctuations of the ordered moment. The bond--operator 
description yields a good account of the magnetization curves and of magnon 
dispersion relations observed by inelastic neutron scattering under applied 
fields, and a variety of experimental predictions for pressure--dependent 
measurements.

\end{abstract} 


\maketitle

\section{Introduction}

Thallium copper chloride\cite{rtst,rsttktmg,roit} presents an insulating,
quantum magnetic system of dimerized $S = 1/2$ Cu$^{2+}$ ions. Inelastic  
neutron scattering (INS) measurements of the elementary magnon  
excitations\cite{rchhfgkm,roktkmm} reveal a strong dispersion in  
all three spatial dimensions indicative of significant interdimer  
interactions. The dispersion minimum gives a spin gap $\Delta_0 = 0.7$meV,  
which is significantly smaller than the antiferromagnetic (AF) dimer  
superexchange parameter $J \approx 5$meV. The corresponding critical field, 
$H_{c} = 5.6$T, makes TlCuCl$_3$ one of the few known inorganic systems 
in which the gap may be closed by application of laboratory magnetic
fields.\cite{rsttktmg} Neutron diffraction measurements at fields $H >  
H_{c}$ revealed that a field--induced AF order in the plane normal to the  
applied field appears simultaneously with the uniform moment.\cite{rtokuokh}  
Recent INS measurements of the magnon spectra in finite fields,\cite{rcrfgkmv} 
including those exceeding $H_{c}$,\cite{rrcfkgvm} have provided dynamical  
information concerning the elementary excitations, in particular the linear 
Goldstone mode,\cite{rrcfgkmwhv} in the phase of field--induced magnetic 
order.  
 
TlCuCl$_3$ (Fig.~1) is one member of a group of related compounds. The 
potassium analog 
KCuCl$_3$\cite{rtst,rsttktmg,rotatstktk,rkttsmnk,rchfgkm,rchhfgkm2,rcrfgkmv}
is similarly dimerized, but has significantly weaker interdimer couplings, 
resulting in a large spin gap of 2.6meV. A further material in the same 
class, NH$_4$CuCl$_3$, has no spin gap and exhibits magnetic order with 
a very small moment, but also shows a complicated low--temperature structure 
which gives rise to magnetization plateaus only at 1/4 and 3/4 of the 
saturation value.\cite{rstktuotmg} While the apparent increase of interdimer 
couplings with anion size may suggest a contribution of the anion to 
superexchange processes, it should be noted that the physical origin of 
the properties of NH$_4$CuCl$_3$ may be rather different from the other 
members.\cite{Matsumoto} Turning from chemical to physical pressure, Tanaka 
{\it et al.}\cite{rtpc} found by magnetization measurements under hydrostatic 
pressure that TlCuCl$_3$ has a pressure--induced magnetically ordered phase,
with a very small critical pressure for the onset of magnetic order, $P_c 
= 2$kbar. Oosawa {\it et al.}\cite{rofokt} have shown very recently by 
elastic neutron scattering measurements under a pressure of 1.48GPa that 
the pressure--induced ordered phase has a strong staggered moment (60\% 
of the saturation value), again reflecting the low value of $P_c$. The 
magnetic Bragg reflections are found at reciprocal lattice points $\bQ 
= (0,0,2\pi)$ (following the notation of Ref.~\onlinecite{rchhfgkm}), 
as in the field--induced ordered phase of \Tl. The aim of the present work 
is to compare and contrast the field-- and pressure--induced ordered phases 
of the system, and to provide a complete description of the static 
magnetization and dynamical excitations at all fields and pressures. 
 
The issue of coupled spin liquids and closing of the spin gap due to  
higher--dimensional superexchange interactions has been investigated in a  
wide variety of systems, many of which were found to be rather close to  
the quantum phase transition (QPT) to an ordered state. The Haldane chains  
CsNiCl$_3$ and RbNiCl$_3$\cite{rtbk} presented an early example where both  
static and dynamic properties could be measured. The coupled, two--leg  
ladder system LaCuO$_{2.5}$\cite{rht} was found to have sufficiently  
strong interladder coupling that magnetic order could set in, whereas the  
coupled plaquette compound CaV$_4$O$_9$\cite{rbg} retains a robust spin  
gap. While numerical simulations of appropriate models have compared the  
thermodynamic properties of the phases on both sides of the  
transition,\cite{rtkzu} theoretical studies of these systems have  
concentrated in particular\cite{ra,raw,rpc,rcm,rhjs,rnr2,retd} on the  
identification of a low--lying but massive longitudinal mode on the  
ordered side of the transition which corresponds to an amplitude fluctuation  
of the ordered moment. Experimental work continues to refine the observations  
of this type of excitation.\cite{rltn} 
 
The first system to be discovered in which the spin gap could be closed  
by application of a magnetic field was the organometallic compound  
(C$_5$H$_{12}$N$_2$)$_2$Cu$_2$Cl$_4$.\cite{rcclpmm} This material, originally  
believed to have a two--leg ladder geometry, undergoes a quantum phase  
transition (QPT) at $H_{c} = 6.6T$, and has been investigated quite  
extensively\cite{rhrbt,rcfjhbhlp,rcjfhlbhp} to characterize the  
thermodynamic properties in the quantum critical regime. The magnetic 
interactions within this system are now thought to have a significantly 
more complex structure,\cite{rsrcyrbfl} whose 3d nature mandates a 
reinterpretation of earlier observations. A further recent discovery,  
(C$_5$H$_{12}$N$_2$)$_2$CuBr$_4$,\cite{rwkmhgmnjbpft} is currently believed  
to be a quasi--1d ladder system, and also shows an almost identical lower  
critical field, $H_c = 6.6T$. A considerable body of theoretical 
work\cite{rhpl,rcg,rm,rnkl,rwy} has analyzed the properties of this system 
by a variety of techniques. A particular focus of recent studies in this 
direction, the majority motivated by experiments on \Tl, has been the 
possibility of describing the ordered phase as a field-- or pressure--driven 
Bose--Einstein condensation (BEC) of the triplet magnon excitations of the 
spin liquid,\cite{rns,rgt,rwh,rnoot} which become massless at the QPT. In  
this formulation the critical exponents of thermodynamic properties such  
as the magnetization in the ordered phase, are specified as a function of  
dimensionality and magnon dispersion, and allow in addition a 
characterization of the quantum critical regime at finite temperature. 
However, questions remain concerning of the appropriateness and 
universality of this description, and we will address these below.

Here we work exclusively at zero temperature to consider the field-- 
and pressure--induced QPTs in TlCuCl$_3$ using the bond--operator 
technique.\cite{rsb} This approach is naturally suited 
to the spin--liquid regime of dimerized systems\cite{rgrs} and has  
previously been developed both for coupled ladders\cite{rnr1} and for  
describing the neighboring magnetically ordered phases\cite{rnr2,rn,rsvb}  
on an equivalent footing. This analytical framework is thus uniquely  
applicable for a consistent discussion of static properties, and in  
particular of dynamical excitations, on both sides of a QPT. In Sec.~II  
we consider the structure of TlCuCl$_3$ and deduce the parameters of the  
effective model for the elementary magnon excitations in zero field.  
Sec.~III contains the bond--operator description of the quantum phases  
and phase transitions in the TlCuCl$_3$ system as a function of the  
field, and as such constitutes the extended version of our previous  
results for magnetizations and magnon excitations.\cite{rmnrs} In Sec.~IV  
we apply the bond--operator formulation to the pressure--induced QPT,  
extending the treatment of Refs.~\onlinecite{rnr2} and \onlinecite{rsvb} 
to discuss in detail the phase and amplitude modes in \Tl, the effects of 
simultaneous pressure and field, and finally the magnetization at finite 
pressure. Sec.~V presents a summary and conclusions.  
 
\section{Magnetic Band Structure}  
 
TlCuCl$_3$ and KCuCl$_3$ appear to have the structure of a quasi--1d system  
of coupled spin chains. However, INS experiments on both systems at zero 
field reveal an elementary triplet excitation with a spin gap and strong 
dispersion in all three reciprocal--space directions. In this section we 
using a model of coupled dimers to provide a qualitative justification for 
the observed results in terms of superexchange interactions within the 
crystal structure. Although the dimer model is different from those of 
Refs.~\onlinecite{rchhfgkm}, \onlinecite{roktkmm}, and \onlinecite{rchfgkm}, 
the results are perforce in close agreement.
 
\begin{figure}[t!]
\psfig{figure=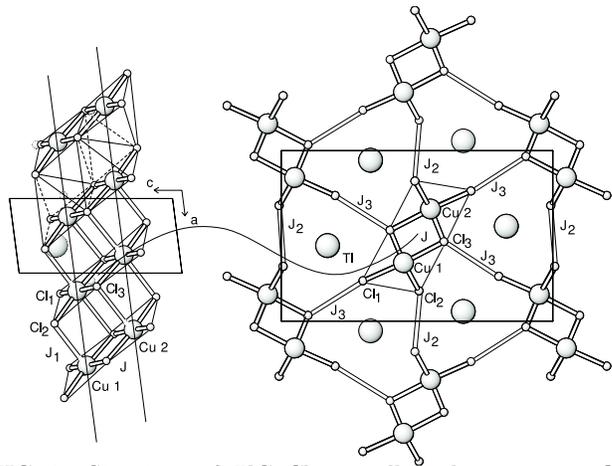,width=8.0cm,angle=0}
\vspace{-0.5cm}
\caption{Structure of TlCuCl$_3$: small circles represent Cl$^-$ ions, 
medium--sized circles Cu$^{2+}$ ions, and large circles Tl$^+$ ions. }
\label{fig:model}
\end{figure}

\subsection{Exchange Pathways}

The magnetic couplings between the spins, which are due to unpaired electrons 
in the Cu $d_{x^2 - y^2}$ orbitals, are mediated by superexchange through the 
$p$ orbitals of the Cl$^-$ ions. Quantitative calculations of superexchange 
interactions, particularly for extended pathways, remain unfortunately beyond 
the scope of current understanding and computer power. Here we employ 
qualitative considerations of geometry and orbital overlap to provide 
plausibility arguments for the relevant model parameters.

Each Cu ion can be considered to be octahedrally coordinated by Cl (Fig.~1). 
The Cu $d_{x^2 - y^2}$ orbitals lie in the basal planes of the octahedra, 
and the most important superexchange interaction, $J$, is that between ions 
in two edge--sharing octahedra. Such a pathway requires exchange through 
quasi--orthogonal orbitals on Cl, and if the environment is truly cubic 
Hund's Rule dictates that the coupling is weakly ferromagnetic (FM). However, 
for a structure distorted away from this high symmetry, as in TlCuCl$_3$, AF 
exchange terms are expected to be stronger, although the edge--sharing  
configuration remains a weak superexchange path on the scale of the linear  
bonds in cuprates. Further superexchange terms in the chains (left side of 
Fig.~1) will of necessity involve pathways coupling to the Cu $d_{3z^2 - 
r^2}$ orbitals, and thus will be considerably weaker than $J$, while 
interchain pathways, which have the form Cu--Cl--(Tl)--Cl--Cu (Fig.~1), 
should also be small. From this consideration one expects a system composed 
of dimers, provided by the edge--sharing octahedra, with weak interdimer 
coupling in all directions.

\begin{figure}[t!] 
\centerline{\psfig{figure=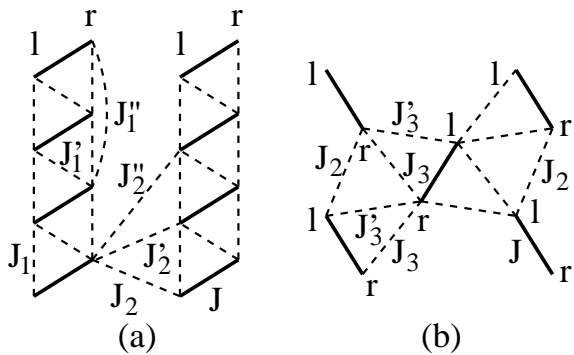,width=7.5cm,angle=0}} 
\vspace{-0.3cm}
\caption{Schematic representation of relevant interdimer couplings in 
XCuCl$_3$: (a) $a$--$c$ plane; (b) $b$--$c$ plane.}  
\end{figure} 

Nearest--neighbor couplings in the chain--like stack of dimer units (oriented  
along ${\hat a}$) are also of the edge--sharing type between octahedra, where  
the edge is from basal plane to apex. Again the possibility arises from the  
geometry that these weak bonds may be FM. Viewing the dimers as ladder rungs,  
the effective ladder leg coupling is denoted by $J_1$, and the asymmetrical  
cross--coupling by $J_{1}^{\prime}$ [Fig.~2(a)]. Finally, the structure of  
the ladder legs is such that a next--neighbor Cu--Cl--Cl--Cu coupling,  
similar to that proposed in CuGeO$_3$, may not be excluded, and this is 
denoted as $J_1^{\prime\prime}$. A significant interchain coupling is 
expected only where the Cu--Cl--Cl--Cu pathway is rather direct, with bond 
angles not deviating widely from 180$^0$. The dimers are oriented primarily 
along the $c$ axis (Fig.~1), where by inspection any overlap is not 
especially direct and may involve parallel paths. Because the dimers are  
also tilted significantly along $\hat{a}$, it is possible that dimers with  
a separation of one or two unit cells in this direction are coupled; let  
$J_2$ denote a dimer coupling along $\hat{c}$ only, $J_{2}^{\prime}$ a  
coupling along $\hat{a} + \hat{c}$ and $J_{2}^{\prime\prime}$ a coupling  
along $\hat{2a} + \hat{c}$ [Fig.~2(a)]. Finally, the shortest interchain  
distance is that between the two inequivalent chains in the unit cell, 
namely $\textstyle{\frac{1}{2}} \hat{b} + \textstyle{\frac{1}{2}} \hat{c}$
(Fig.~1). Here the directionality is evident from the fact that the  
pathway goes through the Cl ions pointing maximally up or down (relative  
to the $a$--axis) from the dimers, indicating that this coupling, $J_3$,  
should correspond to the interdimer separations $\eta \hat{a} +  
{\textstyle \frac{1}{2}} \hat{b} + \textstyle{\frac{1}{2}} \eta \hat{c}$, 
where $\eta = \pm 1$. The possibility remains that the couplings of the 
spin on one dimer to either of those on a neighboring dimer are finite, 
requiring a final superexchange parameter, $J_3^{\prime}$ [Fig.~2(b)]; 
inspection of bond lengths and angles (Fig.~1) suggests that $J_3^{\prime} 
< J_3$. 
 
\subsection{Model} 
 
Having argued that the properties of the system in zero field, including the 
spin gap, are determined largely by the dominant dimer coupling $J$, one may  
construct a model with this ground state using the bond--operator  
technique.\cite{rsb} In this representation the $S = 1/2$ spin degrees of  
freedom on each dimer, or ladder rung, are expressed by the bond operators 
\begin{eqnarray} 
| s \rangle & = & s^{\dag} | 0 \rangle = {\textstyle \frac{1}{\sqrt{2}}} 
 \left( | \! \uparrow \downarrow \rangle  - | \! \downarrow \uparrow \rangle  
\right), \label{ebor} \nonumber \\ | t_x \rangle & = & t_{x}^{\dag} | 0  
\rangle = - {\textstyle \frac{1}{\sqrt{2}}} \left( | \! \uparrow \uparrow  
\rangle  - | \! \downarrow \downarrow \rangle \right), \nonumber \\ | t_y  
\rangle & = & t_{y}^{\dag} | 0 \rangle = {\textstyle \frac{i}{\sqrt{2}}} 
\left( | \! \uparrow \uparrow \rangle  + | \! \downarrow \downarrow \rangle  
\right), \\ | t_z \rangle & = & t_{z}^{\dag} | 0 \rangle = {\textstyle  
\frac{1}{\sqrt{2}}} \left( | \! \uparrow \downarrow \rangle + | \! \downarrow  
\uparrow \rangle \right), \nonumber  
\end{eqnarray} 
where the arrows in each state denote the direction of the two spins, and  
all the states created in Eq.~(\ref{ebor}) have total (dimer) $S_z = 0$.  
From the action of the spin operators ${\bf S}_l$ and ${\bf S}_r$  
(denoting left and right sites of the dimer) on these states one may  
deduce\cite{rsb} the correspondence 
\begin{equation} 
S_{l,r}^{\alpha} = \pm {\textstyle \frac{1}{2}} (s^{\dag}  
t_{\alpha} + t_{\alpha}^{\dag} s) - i \epsilon_{\alpha \beta \gamma}  
t_{\beta}^{\dag} t_{\gamma},
\label{ebot} 
\end{equation} 
where $\alpha,\beta,\gamma \equiv x,y,z$.
The spin commutation relations for ${\bf S}_{l,r}$ are recovered if the  
singlet and triplet operators have bosonic statistics. However, the bond  
operators are constrained by the number of physical states available on  
each dimer, one singlet or one of three triplets, to obey the condition  
\begin{equation} 
s_{i}^{\dag} s_{i} + \sum_{\alpha=x,y,z} t_{i,\alpha}^{\dag} t_{i,\alpha} = 1, 
\label{eboc} 
\end{equation} 
which specifies a hard--core bosonic nature.

With a view to the analysis to be presented in the subsequent sections of 
the system under a finite magnetic field, we introduce here the appropriate 
bond--operator description. The operators in Eq.~(\ref{ebor}) are 
not the eigenstates of an external magnetic field, and so we transform 
these to the operators\cite{rn}
\begin{eqnarray} 
t_x & = & {\textstyle \frac{1}{\sqrt{2}}} ( t_+ + t_-), \cr 
t_y & = & {\textstyle \frac{1}{\sqrt{2}}} i (- t_+ + t_-), \cr 
t_z & = & t_0 , 
\label{eqn:transformation-1} 
\end{eqnarray} 
which create the triplet states 
\begin{eqnarray} 
t_+^\dag |0\rangle & = & - |\uparrow\uparrow\rangle, \cr 
t_0^\dag |0\rangle & = & {\textstyle \frac{1}{\sqrt{2}}} 
        ( |\uparrow\downarrow\rangle + |\uparrow\downarrow\rangle ), \cr 
t_-^\dag |0\rangle & = & |\downarrow\downarrow\rangle. 
\end{eqnarray} 
In this basis the magnetic--field term for a dimer has the diagonal  
representation  
\begin{equation} 
- {\bh \cdot} \left( {\bS}_1 \! + \! {\bS}_2 \right) = i h_z  
( t_x^{\dag} t_y \! - \! t_y^{\dag} t_x ) = h ( t_+^{\dag} t_+ \! -  
\! t_-^{\dag} t_- ) ,  
\label{ebomft} 
\end{equation} 
in which ${\bh} = g \mu_{\rm B} {\bH}$, ensuring that the operators  
$t_{0, \pm 1}^{\dag}$ reproduce the energy levels of the field eigenstates  
$S_z = 0, \pm 1$. Further, the transformation conserves particle number,  
and the local constraint (\ref{eboc}) retains the form
\begin{equation} 
s_i^\dag s_i + \sum_{\alpha = +,0,-} t_{i\alpha}^\dag t_{i\alpha} = 1.
\label{constraint}
\end{equation}
Following Ref.~\onlinecite{rgrs},
transformation to the bond--operator representation yields
\begin{equation}
H = H_0 + H_1 + H_1' + H_1'' + H_2 + H_2' + H_2'' + H_3 + H_3',
\label{ebosh}
\end{equation}
where
\begin{eqnarray}
{\cal H}_0 & \! = \! & \sum_i [ \epsilon_{is} s_i^\dag s_i + 
\sum_{\alpha=+,0,-} \epsilon_{i\alpha} t_{i\alpha}^\dag t_{i\alpha} + 
\mu_i], \nonumber \\
{\cal H}_1 & \! = \! & {\textstyle \frac{1}{2}} J_1 \sum_i
\left[ {\cal H}_{\rm st}(i,i+\ha_1) + {\cal H}_{\rm tt}(i,i+\ha_1) \right], 
\nonumber \\
{\cal H}_1' & \! = \! & {\textstyle \frac{1}{4}} J_1' \sum_i
\left[-{\cal H}_{\rm st}(i,i+\ha_1) + {\cal H}_{\rm tt}(i,i+\ha_1) \right], 
\nonumber \\
{\cal H}_1'' & \! = \! & {\textstyle \frac{1}{2}} J_1'' \sum_i
\left[ {\cal H}_{\rm st}(i,i+\ha_1'') + {\cal H}_{\rm tt}(i,i+\ha_1'') \right],
 \nonumber \\
{\cal H}_2 & \! = \! & {\textstyle \frac{1}{4}} J_2 \sum_i
\left[-{\cal H}_{\rm st}(i,i+\ha_2) + {\cal H}_{\rm tt}(i,i+\ha_2) \right], 
\nonumber \\
{\cal H}_2' & \! = \! & {\textstyle \frac{1}{4}} J_2' \sum_i
\left[-{\cal H}_{\rm st}(i,i+\ha_2') + {\cal H}_{\rm tt}(i,i+\ha_2') \right], 
\nonumber \\
{\cal H}_2'' & \! = \! & {\textstyle \frac{1}{4}} J_2'' \sum_i
\left[-{\cal H}_{\rm st}(i,i+\ha_2'') + {\cal H}_{\rm tt}(i,i+\ha_2'') \right],
 \nonumber \\
{\cal H}_3 & \! = \! & {\textstyle \frac{1}{4}} J_3 \sum_{i,\pm}
            \left[ {\cal H}_{\rm st}(i,i+\ha_{3\pm})
                 + {\cal H}_{\rm tt}(i,i+\ha_{3\pm}) \right], \nonumber \\
{\cal H}_3' & \! = \! & {\textstyle \frac{1}{4}} J_3' \sum_{i,\pm}
            \left[- {\cal H}_{\rm st}(i,i+\ha_{3\pm})
                 + {\cal H}_{\rm tt}(i,i+\ha_{3\pm}) \right]. \nonumber
\end{eqnarray}
In ${\cal H}_0$, $\epsilon_{is}$ and $\epsilon_{i\alpha}$ denote the singlet 
and triplet energies in the presence of an external magnetic field directed 
along the $z$-axis, and $\mu_i$ is a chemical potential emerging from the 
Lagrange multiplier introduced to enforce the constraint (\ref{constraint}) 
on each dimer. The bond Hamiltonian ${\cal H}_{\rm st}(i,j)$ is 
specified by
\begin{equation}
{\cal H}_{\rm st}(i,j) =
   \sum_{\alpha} (t_{i\alpha}^\dag t_{j\alpha} s_j^\dag s_i + 
              t_{i\alpha}^\dag t_{j\bar{\alpha}}^\dag s_i s_j + {\rm H.c.}),
\label{ehsti}
\end{equation}
where $\alpha = +,0,-$ and $\bar{\alpha} = -,0,+$, and describes
singlet--triplet interaction processes: the first two terms in 
Eq.~(\ref{ehsti}) correspond respectively to triplet propagation and 
triplet pair creation. The triplet--triplet interaction term is given by
\begin{eqnarray}
{\cal H}_{\rm tt}(i,j) & = & [t_{j0}^\dag t_{i0}(t_{i+}^\dag
t_{j+}+t_{i-}^\dag t_{j-})+{\rm H.c.}] \label{ehtti} \nonumber \\
& & - [t_{i0}t_{j0}
(t_{i+}^\dag t_{j-}^\dag+t_{i-}^\dag t_{j+}^\dag)+{\rm H.c.}] \\
& & + (t_{i+}^\dag t_{i+} - t_{i-}^\dag t_{i-})
         (t_{j+}^\dag t_{j+} - t_{j-}^\dag t_{j-}), \nonumber
\label{eqn:Htt}
\end{eqnarray}
and encodes the possible triplet scattering processes. As in previous 
analyses\cite{rgrs,rnr2} we have chosen to neglect terms of the form 
$t_{i\alpha}^\dag t_{j\beta}^\dag t_{j\gamma} s_i$: these three--triplet 
interactions vanish identically for a centrosymmetric system,\cite{rsb,rgrs} 
as is the case for \Tl~in the disordered phase, and in the ordered phase their 
contributions may be assumed to be small. Finally, $\ha_i$ specifies the 
bond vectors
\begin{eqnarray}
\ha_1 & = & \ha, \cr
\ha_1'' & = & 2\ha, \cr
\ha_2 & = & \hc, \cr
\ha_2' & = & \ha + \hc, \cr
\ha_2'' & = & 2\ha + \hc, \cr
\ha_{3\pm} & = & \pm \ha + {\textstyle \frac{1}{2}} \hb \pm {\textstyle
\frac{1}{2}} \hc,
\end{eqnarray}
for the superexchange interactions of Fig.~2, where $\ha$, $\hb$, and
$\hc$ are unit vectors for the crystal axes. The prefactor of each term 
in Eq.~(\ref{ebosh}) is determined by the number of bonds between dimers, 
and the sign of ${\cal H}_{\rm st}$ by whether the bond is between 
like ($ll$, $rr$) or unlike spins ($lr$). 

\subsection{Conventional bond--operator formulation}

At zero field, as for all magnetic fields below the critical field $H_{c}$, 
the system is in the quantum disordered regime with a spin gap between the 
singlet and triplet states on each dimer. This situation, for which the 
bond--operator techique is most directly applicable, is represented by 
neglecting the dynamics of the singlet operator and replacing $s_i$ 
everywhere by a $c$--number, ${\overline s}_i$, corresponding to a 
condensate of dimer singlets. Two approaches for the treatment of the 
local constraint (\ref{constraint}) in the disordered phase exist in the 
literature, and we begin with a brief discussion of the conventional 
bond--operator theory,\cite{rsb} in which the chemical potential is 
retained explicitly. Proceeding within a mean--field approximation, the 
operators $s_i$ and the site--dependent chemical potentials $\mu_i$ are 
replaced by uniform, global average values $\langle s_i \rangle = 
\overline{s}$ and $\mu_i = \mu$. These parameters are then determined 
self--consistently from a minimization of the total energy at fixed total 
boson number. The dimer energies $\epsilon_{is}$ and $\epsilon_{i\alpha}$ 
in ${\cal H}_0$ take the values
\begin{eqnarray}
\epsilon_{is} & = & - {\textstyle \frac{3}{4}} J - \mu, \cr
\epsilon_{i\alpha} & = & {\textstyle \frac{1}{4}} J - \alpha h - \mu ,
\end{eqnarray}
and the self--consistent solution follows exactly the treatment of 
Refs.~\onlinecite{rnr1} and~\onlinecite{rnr2}, which also present the 
situation for a system with two dimers per unit cell. 

Quantum fluctuations about this dimer--singlet ground state are contained 
in the triplet operators $t_{i\alpha}^{\dag}$ in Eqs.~(\ref{ebosh}) and 
(\ref{ehsti}), and in reciprocal space are given to quadratic order by the 
Hamiltonian
\begin{eqnarray}
{\cal H} & = & N (- {\textstyle \frac{3}{4}} J \overline{s}^2 
- \mu\overline{s}^2 + \mu) \label{eqn:hamiltonian-disorder} 
\\ & & + \sum_\bsk \left\{ \!\! \sum_{\alpha=+,0,-} \!\! 
\epsilon_{\bsk\alpha} t_{\bsk\alpha}^\dag t_{\bsk\alpha} + 
{\textstyle \frac{1}{2}}(\Lambda_\bsk t_{\bsk 0} t_{-\bsk 0} + {\rm H.c.}) 
\right. \nonumber \\ & & ~~~~~~~~ + \left. \Lambda_\bsk (t_{\bsk +} 
t_{-\bsk -} + t_{\bsk -} t_{-\bsk +} + {\rm H.c.}) \right\}, \nonumber 
\end{eqnarray}
where $N$ is the number of dimers in the system. The dispersive part of 
the magnon energies for the TlCuCl$_3$ system is contained in the expressions
\begin{eqnarray}
\epsilon_{\bsk\alpha} & = & {\textstyle \frac{1}{4}} J - \mu + \Lambda_\bsk 
- \alpha h, \\
\Lambda_\bsk & = & \overline{s}^2 \{ (J_1 - {\textstyle \frac{1}{2}} J_1') 
\cos{k_x} + J_1'' \cos{2k_x} \label{eepsk} \nonumber \\ & & ~~~ 
- {\textstyle \frac{1}{2}} [ J_2 \cos k_z + J_2' \cos(k_x + k_z) 
\nonumber \\ & & ~~~ + J_2'' \cos(2k_x + k_z) ] \nonumber \\ & & ~~~
+ (J_3 - J_3') \cos{(k_x + {\textstyle \frac{1}{2}} k_z )}
\cos{\textstyle \frac{1}{2}} k_y \},
\end{eqnarray}
where $k_z$ takes values in the interval $- 2\pi < k_z \le 2\pi$ in order 
to describe the two dispersion branches corresponding to the two--sublattice 
system. The Brillouin zone lies between $-\pi$ and $\pi$ (in units of the 
inverse lattice spacing) in the other two directions. Diagonalization of 
the Hamiltonian (\ref{eqn:hamiltonian-disorder}) by Bogoliubov 
transformation gives three modes with dispersion relations $E_{\bf k} + h$, 
$E_{\bf k}$, and $E_{\bf k} - h$, where 
\begin{equation}
E_\bsk = \sqrt{\left({\textstyle \frac{1}{4}} J - \mu + \Lambda_\bsk 
\right)^2 - \Lambda_\bsk^2}
\label{eqn:dispersion}
\end{equation}
is the dispersion relation for the field--free system. In the disordered 
regime the triplet modes do not change the form of their dispersion,
and are merely split by the magnetic field due to the Zeeman interaction.

\subsection{Holstein--Primakoff expansion}

The alternative means of treating the global constraint is
to replace the singlet operator by using the constraint itself,
\begin{equation}
s_i = s_i^\dag = \sqrt{ 1 - \frac{1}{N} \sum_{i,\alpha=+,0,-}
                            t_{i\alpha}^\dag t_{i\alpha} },
\label{eqn:global-constraint-1}
\end{equation}
which leads to the simultaneous elimination of the variable $\mu$.
Substitution of Eq.~(\ref{eqn:global-constraint-1}) into the Hamiltonian,
followed by truncation at quadratic order after the expansion of the 
square root, is equivalent to taking ${\overline s} = 1$ and $\mu = 
- {\textstyle \frac{3}{4}} J$ in the formulation of the previous subsection, 
and at higher fields (beyond $H_c$) to specifying that the condensate 
amplitude remains unity. With these values of ${\overline s}$ and 
$\mu$ the results of this approximation have exactly the form of 
Eqs.~(\ref{eqn:hamiltonian-disorder})--(\ref{eqn:dispersion}).

The Holstein--Primakoff approximation thus neglects quantum fluctuations to 
triplet dimers within the singlet condensate, and as such constitutes a 
stronger approximation than the approach of the preceding subsection, in 
which these fluctuations are included in the self-consistent determination 
of $\overline{s}$. The values of ${\overline s}$ obtained for the disordered 
phase in \Tl~and \K~are shown in Table~I, and demonstrate that the triplet 
fluctuations are always small. The maximal error incurred is of order 3\% 
in this regime, while at saturation, where quantum fluctuations are 
completely suppressed, the condensate amplitude is in any case forced to 
unity. In the light of this observation, truncation at quadratic order, 
which is equivalent to the neglect of magnon--magnon interactions in both 
disordered and ordered phases, is also not a source of significant error. 
The superexchange parameters fitted within each of the two approaches 
(Table I) allow a further comparison, and are also in rather close 
agreement, as discussed in detail in the following subsection. Thus the 
Holstein--Primakoff treatment yields semi--quantitative accuracy for 
3d systems such as \Tl~and \K~at low temperatures, and we will use this 
level of approximation in all of the analysis to follow. 

\begin{table}[t!]
\begin{tabular}{|c|c|c|} \hline
& TlCuCl$_3$ & KCuCl$_3$ \\ \hline
$J$/meV $\,$ & $\,$ 5.50 ( 4.87 ) $\,$ & $\,$ 4.22 ( 4.16 ) \\
${\tilde J}_1$/meV $\,$ & $\,$ $-$0.43 ($-$0.41) $\,$ & $\,$ $-$0.42 
($-$0.48) \\
$J_2''$/meV $\,$ & $\,$ 3.16 ( 3.31 ) $\,$ & $\,$ 0.79 ( 0.75 ) \\
${\tilde J}_3$/meV $\,$ & $\,$ 0.91 ( 1.00 ) $\,$ & $\,$ 0.70 
( 0.68 ) \\ \hline
$\Delta_0$/meV $\,$ & $\,$ 0.70 ( 0.71 ) $\,$ & $\,$ 2.60 ( 2.65 ) \\ \hline
${\overline s}$ $\,$ & $\,$ 1.00 ( 0.97 ) $\,$ & $\,$ 1.00 ( 1.00 ) \\
$-\mu/J$ $\,$ & $\,$ 0.75 ( 0.86 ) $\,$ & $\,$ 0.75 ( 0.77 ) \\ \hline
\end{tabular}
\medskip
\caption{Dimer superexchange parameter, $J$, interdimer superexchange
parameters, $J_i$, gap, $\Delta_0$, singlet condensation parameter,
${\overline s}$, and chemical potential, $\mu$, determined from the
bond--operator analysis of TlCuCl$_3$ and KCuCl$_3$. Values to the left 
in each column are obtained from the Holstein--Primakoff formulation of 
Sec.~IID, while values in parenthesis are obtained from the conventional 
bond--operator theory of Sec.~IIC. ${\tilde J}_1$ denotes the combination 
$2 J_1 - J_1'$ and ${\tilde J}_3$ the combination $J_3 - J_3'$.}
\end{table}

\begin{figure}[t!]
\centerline{\psfig{figure=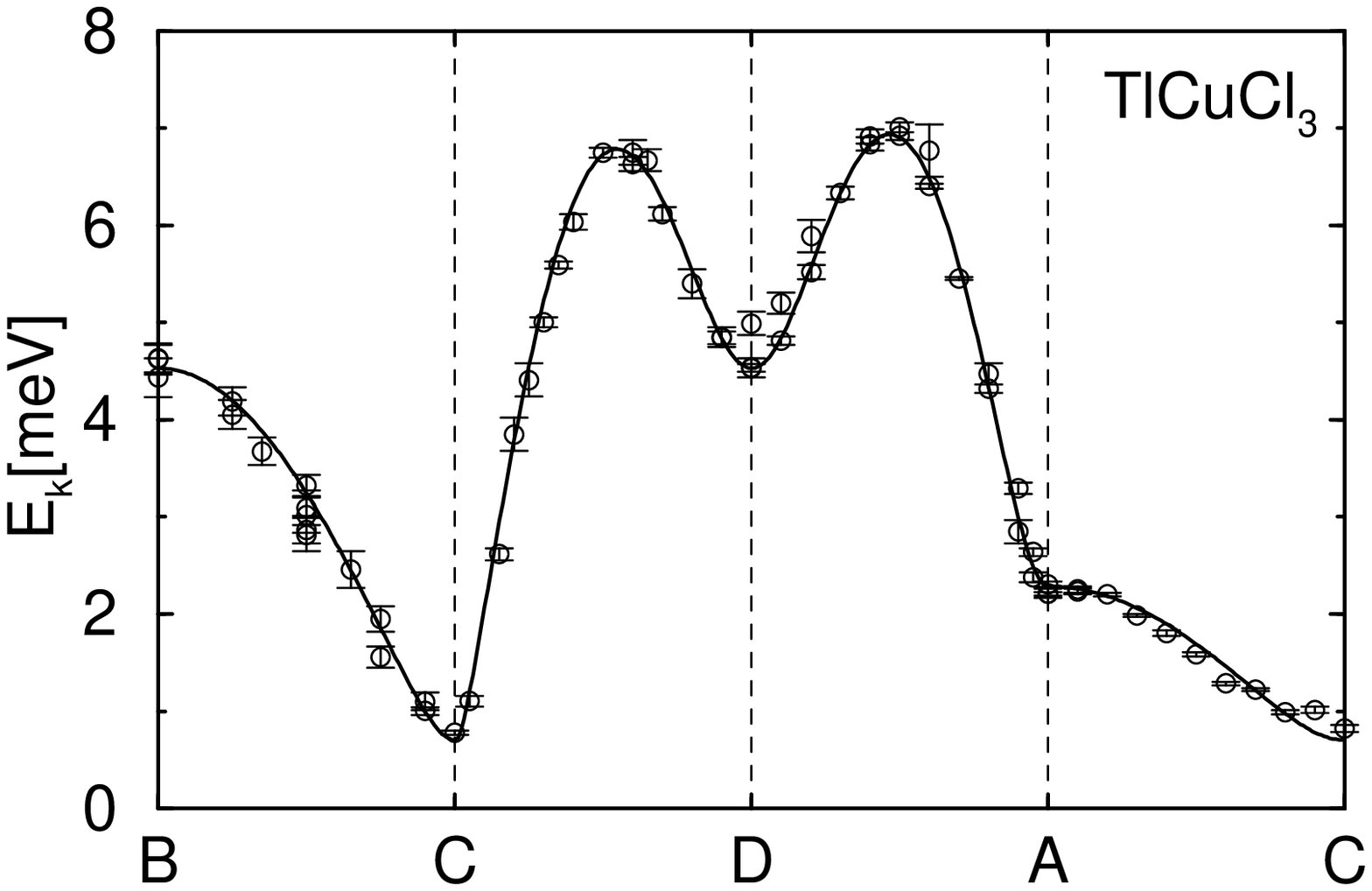,width=7cm,angle=0}}
\centerline{\psfig{figure=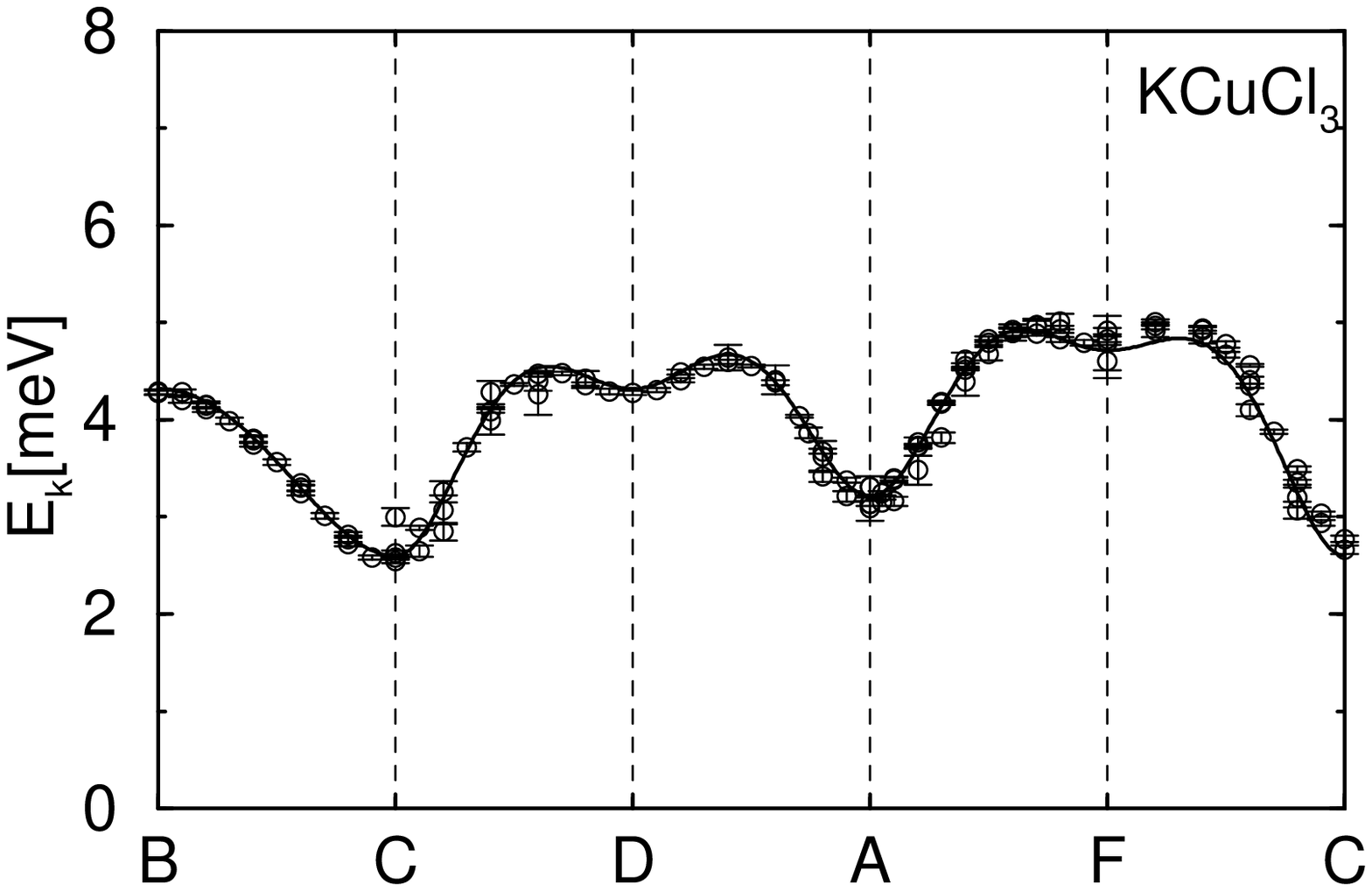,width=7cm,angle=0}}
\caption{(a) Magnon dispersion $E_{\bf k}$ for TlCuCl$_3$. The points 
represent experimental results measured at $T = 1.5$K, taken from 
Ref.~\protect\onlinecite{rchhfgkm}, and the solid lines the fit using the
parameters obtained within the Holstein--Primakoff formulation listed in 
Table I. (b) Magnon dispersion for KCuCl$_3$. Experimental data, measured 
at 5K, is taken from Ref.~\protect\onlinecite{rchfgkm}, and the theoretical 
fit from the Holstein--Primakoff parameters of Table I. Following 
Ref.~\protect\onlinecite{rchhfgkm}, points in reciprocal space are denoted 
by B = (0,2$\pi$,2$\pi$), C = (0,0,2$\pi$), D = (0,0,0), A = ($\pi$,0,0), 
and F = ($\pi$,0,2$\pi$).}
\label{fdt}
\end{figure}

\subsection{Superexchange parameters}

The zero--field dispersion relations (\ref{eqn:dispersion}) deduced from the 
model in Fig.~2 may be compared directly with the experimental results for 
TlCuCl$_3$ and KCuCl$_3$, which we take respectively from 
Refs.~\onlinecite{rchhfgkm} and~\onlinecite{rchfgkm}. The most characteristic 
features of the dispersion data for both systems are that local minima appear 
at both the 0 and $\pi$ points in both $k_x$-- and $k_z$--directions, and 
that the relative depths of these minima in ${\hat k}_x$ (${\hat k}_z$) are 
exchanged between 0 and $\pi$ in ${\hat k}_z$ (${\hat k}_x$). Without 
attempting a systematic optimization procedure it is possible to find 
parameters (Table I) giving a very good account of the data. Fig.~\ref{fdt}(a) 
shows the calculated magnon dispersion for TlCuCl$_3$ with the data of 
Ref.~\onlinecite{rchhfgkm} and Fig.~\ref{fdt}(b) the dispersion for 
KCuCl$_3$ with the data of Ref.~\onlinecite{rchfgkm}. 

Under the assumption that the $\pm \hat{a} + {\textstyle \frac{1}{2}} \hat{b} 
\pm \textstyle{\frac{1}{2}} \hat{c}$ coupling is indeed dominated by $J_3$, 
the signs of the superexchange interactions are consistent with the 
expectation that all intersite parameters $\{J\}$ in Fig.~2 are AF. The 
values for \K~are consistent with the results of Refs.~\onlinecite{rotatstktk},
\onlinecite{rchfgkm}, and \onlinecite{rchhfgkm2}, with discrepancies 
arising from the exact formulation of the dimer description. The leading 
difference between \Tl~and \K~lies in the value of the interladder coupling 
$J_{2}''$, which is larger for \Tl, and is responsible for the 
contrast between wide--band, small--gap and narrow--band, large--gap  
magnon dispersions exemplified by the two systems. We now discuss the  
dispersion relations and superexchange parameters in further detail.  
 
The dominant contribution to the overall dispersion curves comes from  
${\tilde J}_3$, specifically its $\cos k_x \cos {\textstyle \frac{1}{2}} k_z$  
form. Although the magnitude of this coupling is not large, its geometry  
enhances its importance. Separating the $k_x$-- and $k_z$--dependence arising 
from the terms ${\tilde J}_1 \equiv 2 J_1 - J_1'$, $J_1''$, $J_2$, 
$J_2^{\prime}$, and $J_2^{\prime\prime}$ is in principle a complex task, and 
tuning of these parameters would permit a very accurate fit to the measured 
dispersion. For the purposes of the current, unrefined fit we make use of 
the remarkable observation that the magnon energy in direction [1,0,-2] 
(line AC in Fig.~\ref{fdt} for TlCuCl$_3$; a similar observation was made 
for KCuCl$_3$) is small and disperses only weakly and monotonically. In a 
tight--binding model this may be achieved only if the $(k_x,k_z)$ dispersion 
is largely determined by the term $J_2''$. As in Refs.~\onlinecite{rchhfgkm} 
and~\onlinecite{rchfgkm}, we find that a good fit can be achieved to all of 
the dispersion features using only the parameters ${\tilde J}_1$, 
$J_2^{\prime\prime}$ and ${\tilde J}_3$, with contributions 
from the other parameters being negligible. From this procedure it is not 
possible to distinguish between the $J_1$ and $J_1^{\prime}$ contributions 
to ${\tilde J}_1$, but the result that this is negative is robust. This 
implies that the $J_1$ exchange path is closer to the FM coupling regime 
achieved near Cu--Cl--Cu bond angles of 90$^0$ than is the $J_1^{\prime}$ 
path [Fig.~1(a)]. Although it is likely that in this case both values are 
individually rather small, the bond--operator framework provides a simple 
origin for the near--cancelation of two AF interactions to give a small 
net coupling despite the short bond lengths.  

In such a scheme, the exceptionally large $J_2^{\prime\prime}$ required to  
fit the dispersion of TlCuCl$_3$ (Table~I) has no immediate justification  
from qualitative arguments based on superexchange paths. However, it is 
also by no means implausible given the tilted dimer orientation within 
the unit cell (Fig.~2), which results in all of the bonds on this path 
being rather straight. However, because this coupling provides the only 
$\cos 2 k_x$ modulation, and the only $\cos k_z$ modulation, its value is 
determined very accurately by its contribution to the dips away from the 
dispersion maxima towards the zone boundaries in both $k_x$-- and 
$k_z$--directions, primarily in competition with ${\tilde J}_3$. For the 
strongly dimerized, 3d systems represented by both sets of parameters in 
Table~I, these effects may only be due to competition between interactions 
of different periodicities, and there is no possibility of such dispersion 
forms being provided purely by many--body effects of the type expected in 
weakly coupled spin chains.\cite{rbr} Such competition is ensured for the 
assumed dimer coupling geometry by the presence of the pathways within 
${\tilde J}_3$.  

As with ${\tilde J}_1$, the zero--field data alone does not permit the 
possibility of fitting both contributions $J_3$ and $J_3^\prime$ to the 
effective coupling ${\tilde J}_3$. However, because the minimum of the 
magnon bands is located at $(0,0,2\pi)$ in reciprocal space, one may expect 
(on the assumption that all couplings are AF) that $J_1' > 2J_1$, whence 
the negative signs for the effective parameters ${\tilde J}_1$ in Table I, 
and that $J_3 > J_3'$. One attempt has been made\cite{roktkmm} to extract 
these parameters for \Tl~using a cluster expansion theory,\cite{rmm} which 
returns the intersite couplings (valid for a formulation different from that 
of Sec.~II) $J_1 = 0.34$meV, $J_1' = 1.70$meV, $J_3 = 0.91$meV, and $J_3' = 
- 0.57$meV. However, for the present purposes, namely a focus on the ordered 
phases of the system under applied field and pressure, we choose to avoid the 
large uncertainties inherent in these estimates and proceed simply with the 
effective quantities ${\tilde J}_1$ and ${\tilde J}_3$, determined at 
zero field, in place of ${\textstyle \frac{1}{2}} J_1$ and $J_3$ in 
Eq.~(\ref{ebosh}), with ${\cal H}_1$ and ${\cal H}_3'$ set to zero. This 
approximation is in fact insufficient for a full treatment of the ordered 
phase, where the coefficients of the terms ${\cal H}_{\rm tt}$ have the 
same sign, and should no longer be necessary when reliable high--field 
measurements become available. The resulting small deviations from 
quantitative accuracy which are found in reproducing the high--field 
magnetization (Sec.~IIIB) imply that $J_1'$ and $J_3$ are, respectively, 
significantly larger than $J_1$ and $J_3'$. We have thus extracted only 
four exchange constants (Table I) as the minimal model required to describe 
\Tl~and \K, and have shown in Fig.~\ref{fdt} that this set provides a 
remarkably good account of the magnon dispersion relations in both systems. 
In the following sections we therefore employ this model with the parameters, 
$(J, {\tilde J}_1, J_2'', {\tilde J}_3)$, and within the Holstein--Primakoff 
approximation (Sec.~IID), to study field-- and pressure--induced magnetic 
ordering in \Tl~and \K.

We comment here that the XCuCl$_3$ structure is not magnetically frustrated: 
when following the dominant interactions no triangles of spins are found.  
The modulation of the dispersion relations arises rather from the simple  
fact that components of different periodicities may contribute as a  
consequence of the complex unit cell and coupling geometry. The lack of  
frustration is a very important feature for both the appearance of  
staggered magnetic order in an applied field and for the success of the  
bond--operator description.  
 
With regard to a comparison of the parameters in Table I with the results of 
other approaches, these are indeed rather similar to the values given in  
Refs.~\onlinecite{rchhfgkm} [differences are due to the fact that the
experimental fitting formula is linear, whereas Eq.~(\ref{eqn:dispersion}) 
has a square--root form] and \onlinecite{roktkmm}. The bond--operator 
formulation has the advantage of providing a direct justification 
for AF interdimer bonds $J_2^{\prime\prime}$ and ${\tilde J}_3$. The opposite 
sign of the $J_2^{\prime\prime}$ term found by the alternative technique is  
rather harder to justify on the grounds of magnitude and orbital overlap  
geometry. The differences between the results of Table I and those shown in 
Table I of Ref.~\onlinecite{rmnrs} are due solely to the conventions chosen 
for the bond--operator analysis. That the magnon excitations are so well 
described by such a  
simple model indicates {\it a posteriori} that the qualitative assumptions  
made in Sec.~IIA concerning the relevant exchange paths are appropriate.  
The increase in interdimer coupling strengths observed from KCuCl$_3$ to 
TlCuCl$_3$, as the anion size increases, may suggest that the role of the 
anion in determining the superexchange parameters may depend more on its 
charge distribution (active participation in superexchange) than merely 
on its spacing effect.  
 
In summary, TlCuCl$_3$ and KCuCl$_3$ are well described by simple models  
of dimerized Cu spins coupled by relatively weak superexchange interactions  
in all three spatial dimensions. This model may be used not only for magnon  
dispersion relations measured by INS, but also for the intensities of the  
elementary excitations, which are dominated by the dimer form factor. With  
these zero--field results we turn now to a treatment of the excitations in  
magnetic fields $H > H_{c}$ sufficient to close the spin gap in TlCuCl$_3$.  
In Sec.~IV we will return to these considerations to justify our treatment  
of the superexchange integrals under applied hydrostatic pressure.   

\section{Field--induced order}

In the previous section we have established a minimal model within the
bond--operator framework which describes the ground state and magnetic
excitations of TlCuCl$_3$ and KCuCl$_3$ at zero magnetic field. In this 
section we will apply the same formalism to consider the evolution of 
the system under an applied field, with particular focus on the excitations 
of TlCuCl$_3$ in the experimentally accessible intermediate--field regime 
$H > H_{c}$, where the lowest-lying magnon mode is gapless.

\begin{figure}[t!]
\begin{center}
\includegraphics[width=6.0cm]{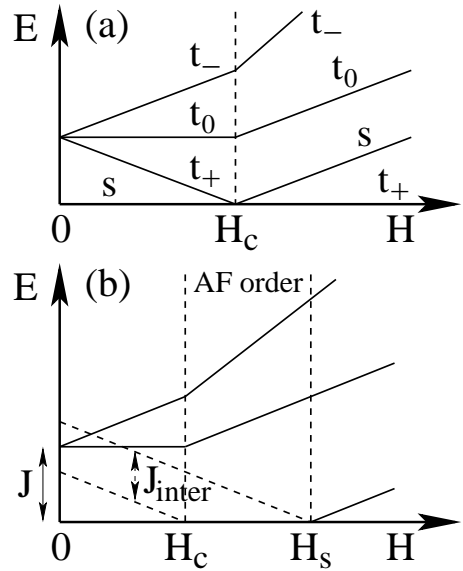}
\end{center}
\vspace{-0.5cm}
\caption{Representation of field--induced order for (a) an isolated dimer
and (b) interacting dimers with interdimer coupling $J_{\rm inter}$.}
\label{fig:summary-field}
\end{figure}

The bond--operator formulation for ground and excited states in magnetically 
ordered phases was first explored for systems where the order arises as a 
result of increasing the superexchange interactions between low--dimensional 
subsystems with a spin gap.\cite{rnr2} In this case, all three magnon modes 
in the disordered phase remain degenerate as the coupling rises and the spin 
gap is closed, until at the quantum critical point (QCP) the three magnons 
become massless at the band minimum (magnetic Bragg point). On further 
increase of the coupling into the ordered phase, two of the magnons remain 
massless, corresponding to spin waves, while the third acquires a mass 
which increases on moving away from the QCP. This last excitation has the 
interpretation of a longitudinal mode, and corresponds to fluctuations in 
the amplitude of the order parameter, which are soft close to the QCP. This 
situation will be discussed further in Sec.~IV in the context of the 
pressure--induced QPT in TlCuCl$_3$. 

The situation in a magnetic field is rather different, and has been 
considered using bond operators for the isolated spin ladder,\cite{rn} 
for a pair of coupled planes,\cite{rsvb} and for TlCuCl$_3$ in 
Ref.~\onlinecite{rmnrs}. A simple understanding may be obtained by 
considering a single, isolated dimer with AF coupling $J$: at zero field 
the ground state is a singlet, and the three--fold degenerate triplet 
excitations at energy $\Delta = J$ separate in an applied field due to the  
Zeeman interaction, as shown schematically in Fig.~\ref{fig:summary-field}(a). 
At a critical field, given by $h_c \equiv \Delta$, the energy of the lowest  
triplet is reduced to zero, and the crossing of levels changes the ground  
state from singlet to triplet. Above $H_c$, the ground state is of pure  
triplet nature and the lowest--lying excitation is a singlet. For a set of  
dimers with mutual AF interactions [Fig.~\ref{fig:summary-field}(b)], 
the zero--field triplet excitation modes are dispersive in reciprocal  
space, and the critical field $H_{c}$ is defined by the value required to 
close the zero--field gap, at which point a single mode becomes massless.

The ground state of the ordered phase has the two essential properties 
that is it possesses a field--induced staggered moment and has a massless 
excitation corresponding to the transverse fluctuations of this moment. 
The appropriate description of this phase is a linear combination with the 
singlet of the lowest-- and highest--lying triplet states,\cite{rmnrs} 
which corresponds to the choice of a mean field for the staggered 
moment, whence the transformation in Holstein--Primakoff formulation 
returns the excitations of the ordered phase. We note here that the 
conventional bond--operator treatment (Sec.~IIC) may lose consistency 
in the description of these excitations because higher--order interactions 
between triplets are contained in an uncontrolled manner. The coefficients 
of the ground--state singlet--triplet admixture change continuously with 
the applied field until the upper critical field $H_{s}$, where all spins 
are aligned. Beyond $H_{s}$, the ground state has pure triplet character 
(Fig.~\ref{fig:summary-field}) and three excited modes of the strong--field 
regime are all massive, as expected for a field--aligned antiferromagnet.

\subsection{Disordered phase}

\begin{figure}[t!]
\begin{center}
\includegraphics[width=7cm]{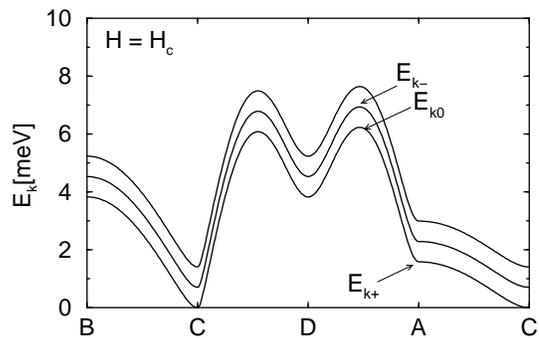}
\end{center}
\vspace{-0.5cm}
\caption{Magnon dispersion for \Tl~at $H = H_c$ (= 5.6T),
using $g = 2.16$ for the $g$--factor of \Tl.}
\label{fig:dispersion-hc}
\end{figure}

Henceforth we adopt the notation $J_1$ for ${\tilde J}_1$, $J_2$ for $J_2''$, 
and $J_3$ for ${\tilde J}_3$. As in Sec.~IIC, the dispersion relations for 
the triplet magnons of the disordered phase are given by
\begin{eqnarray}
E_{\bsk\alpha} &=& \sqrt{J^2 + 2J\epsilon_\bsk} -\alpha h, \\
\epsilon_\bsk & = & - {\textstyle \frac{1}{2}} J_1 \cos{k_x} -
{\textstyle \frac{1}{2}} J_2 \cos{(2k_x + k_z)} \label{eepsk2} \nonumber \\
& & + J_3 \cos{(k_x + {\textstyle \frac{1}{2}} k_z )}
\cos{\textstyle \frac{1}{2}} k_y ,
\label{eqn:dispersion0}
\end{eqnarray}
and are shown in Fig.~\ref{fig:dispersion-hc} for the critical field $h_c$, 
although a qualitatively identical picture is obtained for all fields $h 
\le h_c$. We emphasize that the lowest mode retains its quadratic shape at 
the band minimum ({\it cf.} Sec.~IV) even at $h_c$, while beyond this, as
we will demonstrate below, an additional linear component arises. 

The nature of the three magnon modes in the disordered phase may be 
understood as follows. The wave function of the lowest triplet mode, 
$E_{\bsk +}$, can be approximated by a linear combination of singlets 
and triplets on each dimer bond,  
\begin{equation} 
\psi_i \sim {\textstyle \frac{1}{\sqrt{2}}} u (|\uparrow \downarrow \rangle 
 - |\uparrow \downarrow \rangle) - v e^{i (\bsk\cdot\bsr_i - E_{\bsk +} t)}
|\uparrow \uparrow \rangle, 
\label{eqn:wavefunction-1} 
\end{equation} 
where $u$ is a coefficient of order unity and $v$ is small and real. 
The expectation values of the spin operator components at a given dimer  
$i$ in the state (\ref{eqn:wavefunction-1}) are
\begin{eqnarray} 
\langle S_{l,x} \rangle_i & = & -\langle S_{r,x} \rangle_i 
  \sim {\textstyle \frac{1}{2}} u v \cos(\bk\cdot\br_i \! - \! E_{\bsk +} t), 
\label{eqn:spin} \nonumber \\ 
\langle S_{l,y} \rangle_i & = & -\langle S_{r,y} \rangle_i 
  \sim - {\textstyle \frac{1}{2}} u v \sin(\bk\cdot\br_i \! - \! E_{\bsk 
  +} t) ,\\ 
\langle S_{l,z} \rangle_i & = & \langle S_{r,z} \rangle_i \sim  
{\textstyle \frac{1}{2}} v^2, \nonumber
\end{eqnarray} 
from which it is evident that the mode possesses a very small, uniform,  
static magnetic moment parallel to the field, and thus gains Zeeman energy. 
Perpendicular to the field there is also a finite magnetic moment, 
which fluctuates in both space and time with characteristic wavevector  
$\bk$ and energy $E_{\bsk +}$, and is staggered (spins $l$ and $r$  
oppositely aligned) because of the AF intradimer coupling $J$. Similar 
considerations for the mode $E_{\bsk -}$ reveal that the uniform magnetic 
moment is directed antiparallel to the field, leading to a higher Zeeman 
energy. Finally, the $E_{\bsk 0}$ mode has no magnetization perpendicular 
to the field, although it does possess a moment parallel to the field with 
modulation wavevector $\bk$, and so its energy does not change with field. 
For $H < H_c$, the energies of these modes are positive, meaning that the 
spin structure associated with the magnons is very weak and fluctuating. 
We stress that the sum of the magnetization components remains zero in the 
disordered phases, which can be considered as a consequence of quantum 
fluctuations which act to restore the $O(3)$ symmetry of the field--free 
system.

We note again that an increase in the applied field leads to a shift in  
mode energies without changing the shape of their zero--field dispersion,  
until the lowest mode ($\alpha = +$) becomes soft at $\bQ=(0,0,2\pi)$.  
This determines both the critical field,
\begin{equation}
h_c = g \muB H_c = \sqrt{J^2 - J(J_1 + J_2 + 2J_3)} ,
\label{eqn:hc}
\end{equation}
and the spin structure of the lowest--lying mode, given by $\bk = \bQ$, 
which corresponds to the static spin configuration in the field--induced 
ordered phase. One may conclude that a \Tl~system of classical spins would 
favor AF order to minimize simultaneously all intersite interaction terms. 
By contrast, for the quantum system the strong AF intradimer interaction, 
$J$, favors dimer singlet formation, and the AF spin structure cannot be 
the ground state for $H < H_c$. However, the incipient, long--ranged AF 
spin order is reflected in the structure
of the lowest--lying magnon excitation, which is minimal at $\bk = \bQ$.
As the magnetic field is increased, the lowest mode is lowered
progressively until $H_c$, where dimer singlet formation is no longer
sufficient to cause an ordered phase, the symmetry of the system is
lowered to $O(2)$, and the onset of real uniform and staggered
magnetization components is observed. For the magnon excitations in
the ordered regime, the instability is signaled by the lowest mode
becoming negative within the formulation of this subsection. We turn
now to a consistent description of all of these phenomena in the
ordered phase ($H > H_c$).

\subsection{Ordered phase ($H_c\le H \le H_s$)}

\subsubsection*{Magnon excitations}

In the intermediate--field regime, the ground state of each dimer can be 
considered as a partially polarized FM configuration. It is important to 
note that there is no explicit AF component in the dimer ground state, and 
that the ordering emerges only from closing of the gap to the lowest magnon 
branch at $\bk = \bQ$. In the bond--operator formulation, this ordered
ground state is represented\cite{rsvb,rmnrs} in the basis of
Eq.~(\ref{eqn:transformation-1}) by finite expectation values
${\overline s}$ , ${\overline t_+}$, and ${\overline t_-}$ of these
singlet and triplet operators. The component of the highest--lying triplet
mode in the ground--state condensate may appear counterintuitive, and was
neglected in a number of approximate treatments.\cite{rm,rn} However, the
presence of terms of the form $t_{i+}^{\dag} t_{j-}^{\dag} s_j s_i$
in the transformed Hamiltonian of Eq.~(\ref{ehsti}) makes clear that a 
finite component of this state is required in the consistent condensate.

Following the treatment of the preceding sections, the linear combination of 
operators appropriate for the ground state mode is taken as the condensate, 
and the orthogonal mode containing a singlet component describes its gapless 
phase fluctuations. For a full description of the ordered regime we perform 
the transformation\cite{rsvb}
\begin{eqnarray}
a_i & = & u s_i + v ( f e^{i \bsQ \cdot \br_i} t_{i+} + g e^{i \bsQ \cdot
\br_i }t_{i-} ), \label{eqn:b} \nonumber \\
b_{i+} & = & u ( f t_{i+} + g t_{i-} ) - v e^{i \bsQ \cdot \br_i} s_{i},
\nonumber \\ b_{i0} & = & t_{i0}, \\
b_{i-} & = & f t_{i-} - g t_{i+} , \nonumber
\end{eqnarray}
in which the $\bk$--independent coefficients $u$, $v$, $f$, and $g$ arise  
from successive unitary transformations. The conditions $u^2 + v^2 = 1$  
and $f^2 + g^2 = 1$ dictate that $u = \cos\theta$, $v = \sin\theta$, $f = 
\cos\phi$, and $g = \sin\phi$, with $\theta$ and $\phi$ the two independent  
variables required to specify the condensate.  
 
As above, the operator $a_i$ is treated as a uniformly condensed  
mean--field parameter, 
\begin{equation} 
a_i \longrightarrow {\overline a} . 
\label{ea}
\end{equation} 
Because the two unitary transformations conserve particle number,  
the global constraint may be reduced in the Holstein--Primakoff  
representation
to 
\begin{equation} 
{\overline a} = \sqrt{1 - \frac{1}{N} \sum_{\bsk,\alpha=+,0,-} 
b_{\bsk\alpha}^\dag b_{\bsk\alpha}}. 
\label{eqn:global-constraint-3} 
\end{equation} 
We emphasize again the participation of the highest triplet mode ($t_{-}$) 
in the condensate, as a consequence of the triplet pair--creation  
processes in ${\cal H}_{\rm st}$ (\ref{ehsti}). The linear combination of 
singlet and triplets in the condensate ${\overline a}$ yields a staggered 
magnetization perpendicular to the field, whose order is given by the wave 
vector $\bQ$. 
 
\begin{figure}[t!]
\begin{center}
\includegraphics[width=3.7cm]{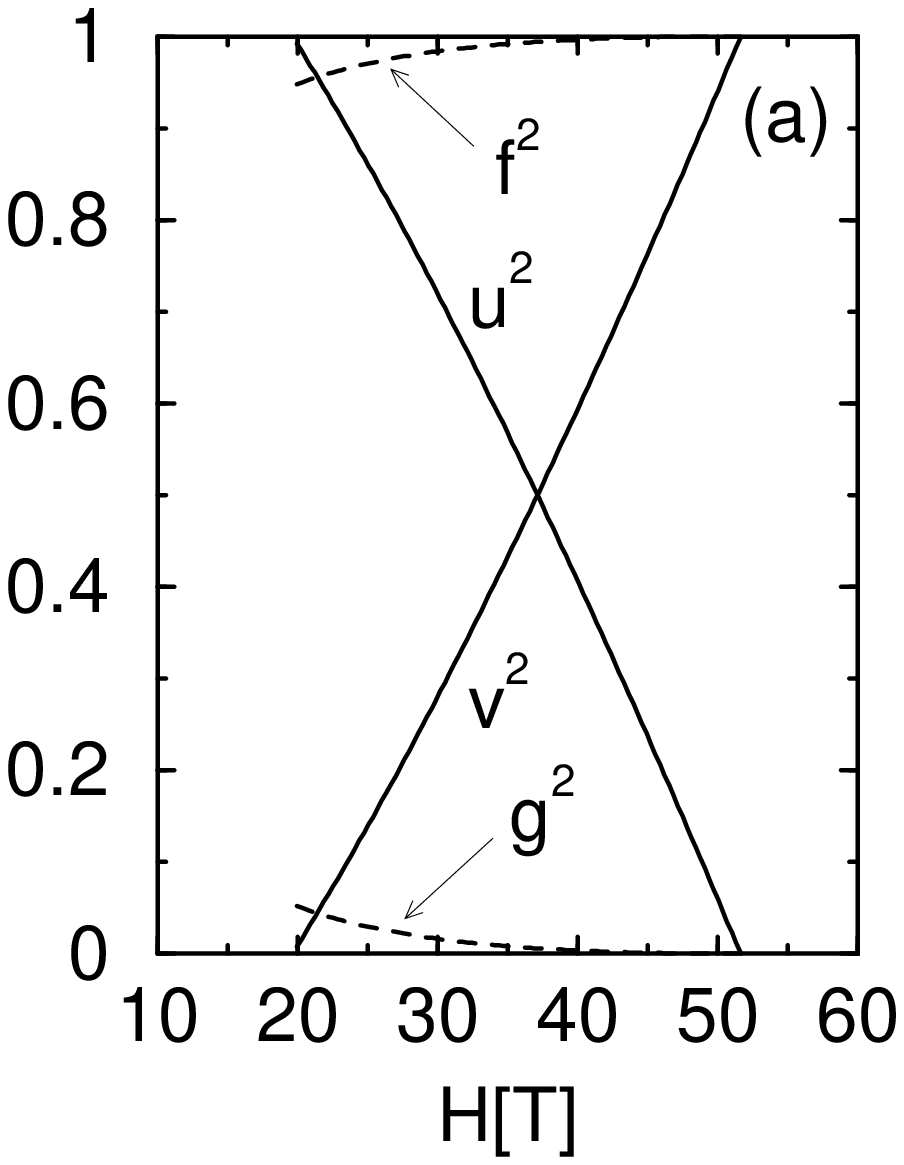}
\includegraphics[width=3.8cm]{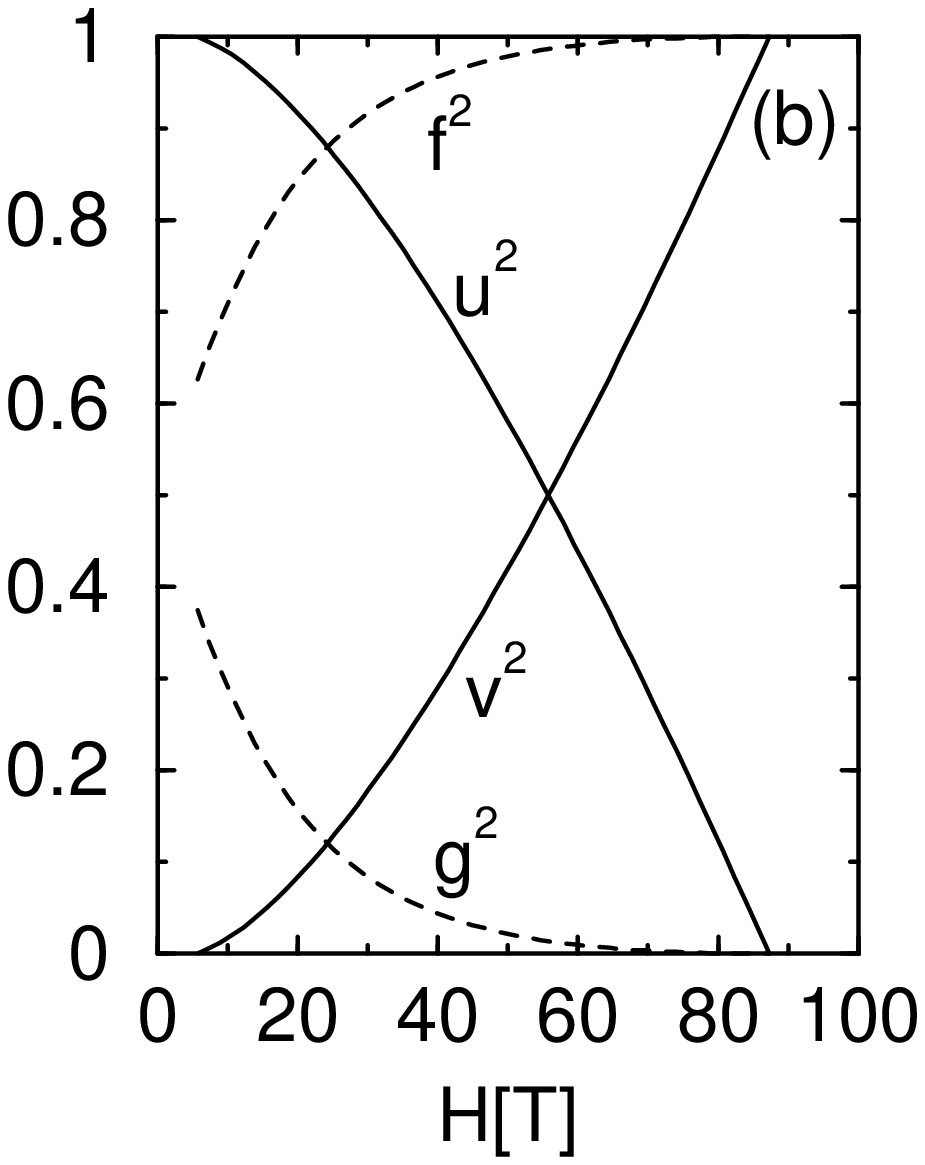}
\end{center}
\vspace{-0.8cm}
\caption{Magnetic--field dependence of the coefficients $u^2$, $v^2$, $f^2$, 
and $g^2$, for (a) \K~and (b) \Tl.}
\label{fig:u}
\end{figure}

Using the transformed operators (\ref{eqn:b}), and with  
replacement of the condensate parameter using the global constraint of 
Eq.~(\ref{eqn:global-constraint-3}), the Hamiltonian (\ref{ebosh})
is reexpressed up quadratic order in the $b$ operators in the form  
\begin{equation} 
{\cal H} = O(b^0) + O(b^1) + O(b^2),  
\label{eqn:hamiltonian-order} 
\end{equation} 
in which the classical part at zeroth order is given by  
\begin{eqnarray} 
O(b^0) & = & N \bigl\{ (\epsilon_+ f^2 + \epsilon_- g^2) v^2 \label{eob0} \\
&&- \epsilon_\bsQ [- u^2 v^2 (1 + 2fg)
    + {\textstyle \frac{1}{2}} v^4(f^2 - g^2)^2]\bigr\}, \nonumber
\end{eqnarray}
where
\begin{eqnarray}
\epsilon_\alpha & = & J - \alpha h,~~~(\alpha=\pm) \\
\epsilon_\bsQ & = & - ({\textstyle \frac{1}{2}} J_1 + {\textstyle 
\frac{1}{2}} J_2 + J_3).
\end{eqnarray}
The terms at linear order may be written as 
\begin{eqnarray} 
O(b^1) & = & u v (a_0^\dag b_{\bsQ +} + {\rm H.c.} ) \{ \epsilon_+ f^2
+ \epsilon_- g^2 \label{eob1} \nonumber \\
 & & + \epsilon_\bsQ  
[(u^2 - v^2) (1 + 2fg) - v^2 (f^2 - g^2)^2]\} \nonumber \\ 
& & + u v (a_0^\dag b_{\bsQ -} + {\rm H.c.} ) \\ 
& & \times [(-\epsilon_+ + \epsilon_-) f g + \epsilon_\bsQ (u^2  
+ 2 f g v^2)(f^2 - g^2)], \nonumber
\end{eqnarray} 
and the requirement that this sum be zero fixes the parameter  
choice $(\theta,\phi)$, which also minimizes $O(b^0)$. The equations  
determining the two unitary transformations are then given by 
\begin{eqnarray} 
\epsilon_+ f^2 + \epsilon_- g^2 & \! = \! & -\epsilon_\bsQ [(u^2 \! - \! 
v^2)(1 \! + \! 2 f g) - v^2 (f^2 \! - \! g^2)^2], \label{eqn:uvfg-fix} 
\nonumber \\ (-\epsilon_+ + \epsilon_-) f g & \! = \! & -\epsilon_\bsQ 
(u^2 \! + \! 2 f g v^2)(f^2 \! - \! g^2). 
\end{eqnarray} 
The field dependence of the coefficients $u^2$, $v^2$, $f^2$, and $g^2$ is 
shown in Fig.~\ref{fig:u} for the parameters of both \K~and \Tl.
Below the critical field ($H < H_c$), the ground state is a  
pure singlet condensate specified by $u = 1$, $v = 0$, $f = 1$,  
and $g = 0$, {\it i.e.}~by the angles ($\theta,\phi$) = (0,0). Precisely  
at the critical field, where $u = 1$ and $v = 0$ ($\theta = 0$), $g$ takes 
on a finite value while $f < 1$. In the intermediate regime, $\theta$ then
changes continuously from 0 to $\pi/2$. Except for the parameters of 
\Tl~and for fields close to $H_c$, $\phi$ remains small, reflecting the fact
that mixing of the upper and lower triplet modes is not generally strong.
Finally, above the saturation field ($H \ge H_s$), the ground state becomes 
a pure condensate of the lowest triplet, which is described by $u = 0$, 
$v = 1$, $f = 1$, and $g = 0$, or by $(\theta,\phi) = (\pi/2,0)$.
The critical field is determined exactly as
\begin{equation}
h_c = g \muB H_c = \sqrt{J^2 - J (J_1 + J_2 + 2 J_3)},
\label{ecf}
\end{equation}
which coincides with the soft--mode condition (\ref{eqn:hc}) in the
low--field regime. For \Tl~(\K), the parameters of Table~I give a
critical field $H_c = 5.6$T ($19.6$T), consistent with the measured
value.\cite{rsttktmg} The saturation field is determined by the
condition
\begin{equation}
h_s = g\muB H_s = J + J_1 + J_2 + 2 J_3
\label{esf}
\end{equation}
required to overcome all of the AF bonds to a single spin.

The spin excitations in the ordered phase are described by the quadratic
term in Eq.~(\ref{eqn:hamiltonian-order}),
\begin{equation}
O(b^2) = {\cal H}_{0} + {\cal H}_\pm,
\label{eob2}
\end{equation}
where
\begin{equation}
{\cal H}_0 = \sum_\bsk [\epsilon_{\bsk 0} b_{\bsk 0}^\dag b_{\bsk 0}
 + {\textstyle \frac{1}{2}} (\Delta_{\bsk 0}^\dag b_{\bsk 0} b_{-\bsk 0}
+ {\rm H.c.})]
\label{eob2h0}
\end{equation}
and
\begin{eqnarray}
{\cal H}_\pm & = & \sum_\bsk \left[ \epsilon_{\bsk +} b_{\bsk +}^\dag
b_{\bsk +} + \epsilon_{\bsk -} b_{\bsk -}^\dag b_{\bsk -} \right.
\label{eob2hpm} \nonumber \\
& & ~~~~~~~ + (\epsilon_{\bsk \pm} b_{\bsk +}^\dag b_{\bsk -} + {\rm H.c.})
\nonumber \\ & & ~~~~~~~ + {\textstyle \frac{1}{2}} (\Delta_{\bsk +}^\dag
b_{\bsk +} b_{-\bsk +} + {\rm H.c.}) \nonumber \\
& & ~~~~~~~ + {\textstyle \frac{1}{2}} (\Delta_{\bsk -}^\dag b_{\bsk -}
b_{-\bsk -} + {\rm H.c.}) \nonumber \\ & & ~~~~~~~ \left.
+ (\Delta_{\bsk \pm}^\dag b_{\bsk +} b_{-\bsk -} + {\rm H.c.}) \right].
\end{eqnarray} 
For continuity of presentation, the expressions for $\epsilon_{\bsk i}$  
and $\Delta_{\bsk i}$ are given in the Appendix [Eqs.~(\ref{eaope}) and 
(\ref{eaopd})]. We note that on setting $(\theta,\phi) = (0,0)$ the 
Hamiltonian [(\ref{eob0}) and (\ref{eob2})] reduces to that of the 
disordered phase given in Eq.~(\ref{eqn:hamiltonian-disorder}) with 
${\overline s} = 1$ and $\mu = - {\textstyle \frac{3}{4}} J$.

\begin{figure}[t!]
\begin{center}
\includegraphics[width=7cm]{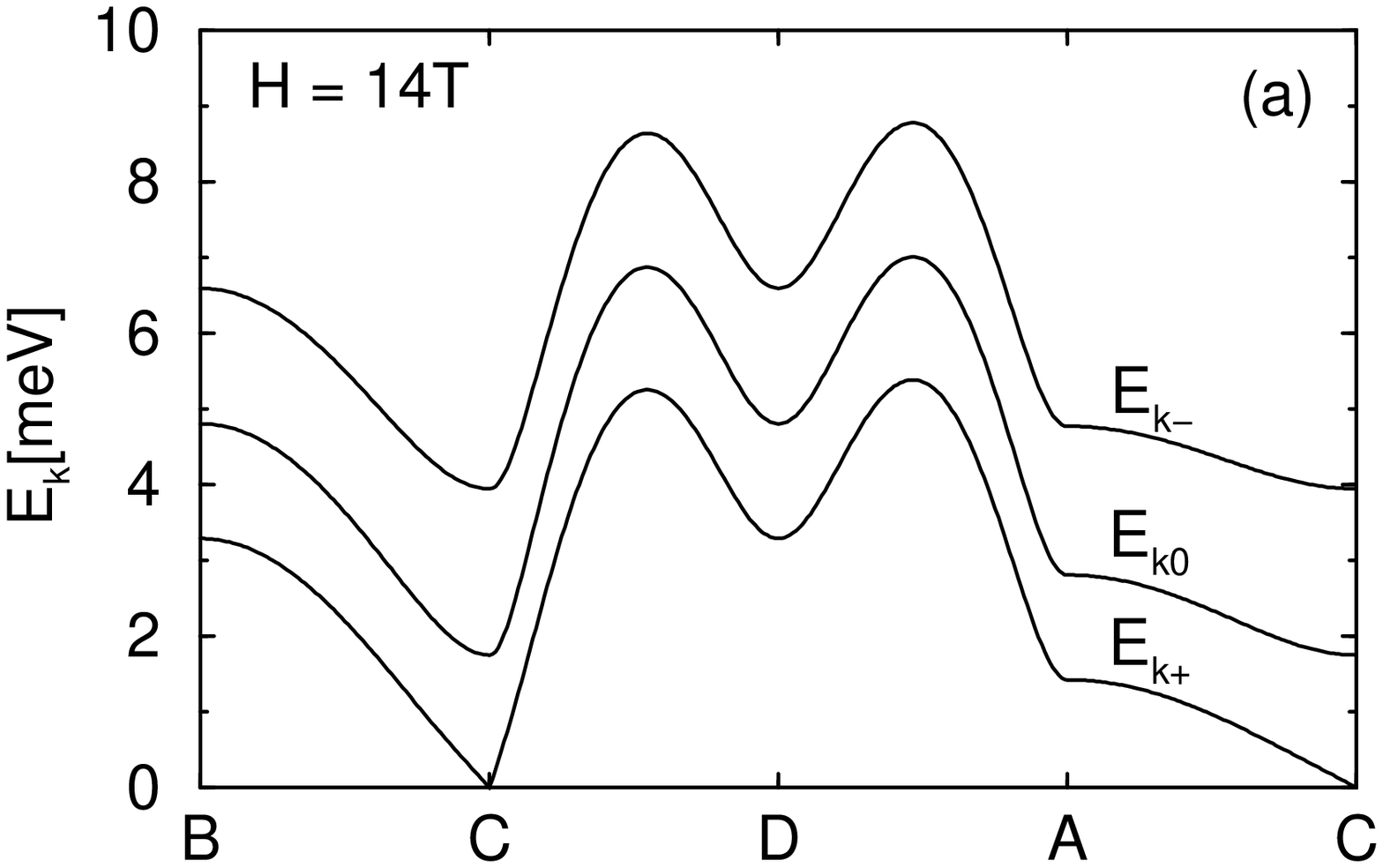}
\includegraphics[width=7cm]{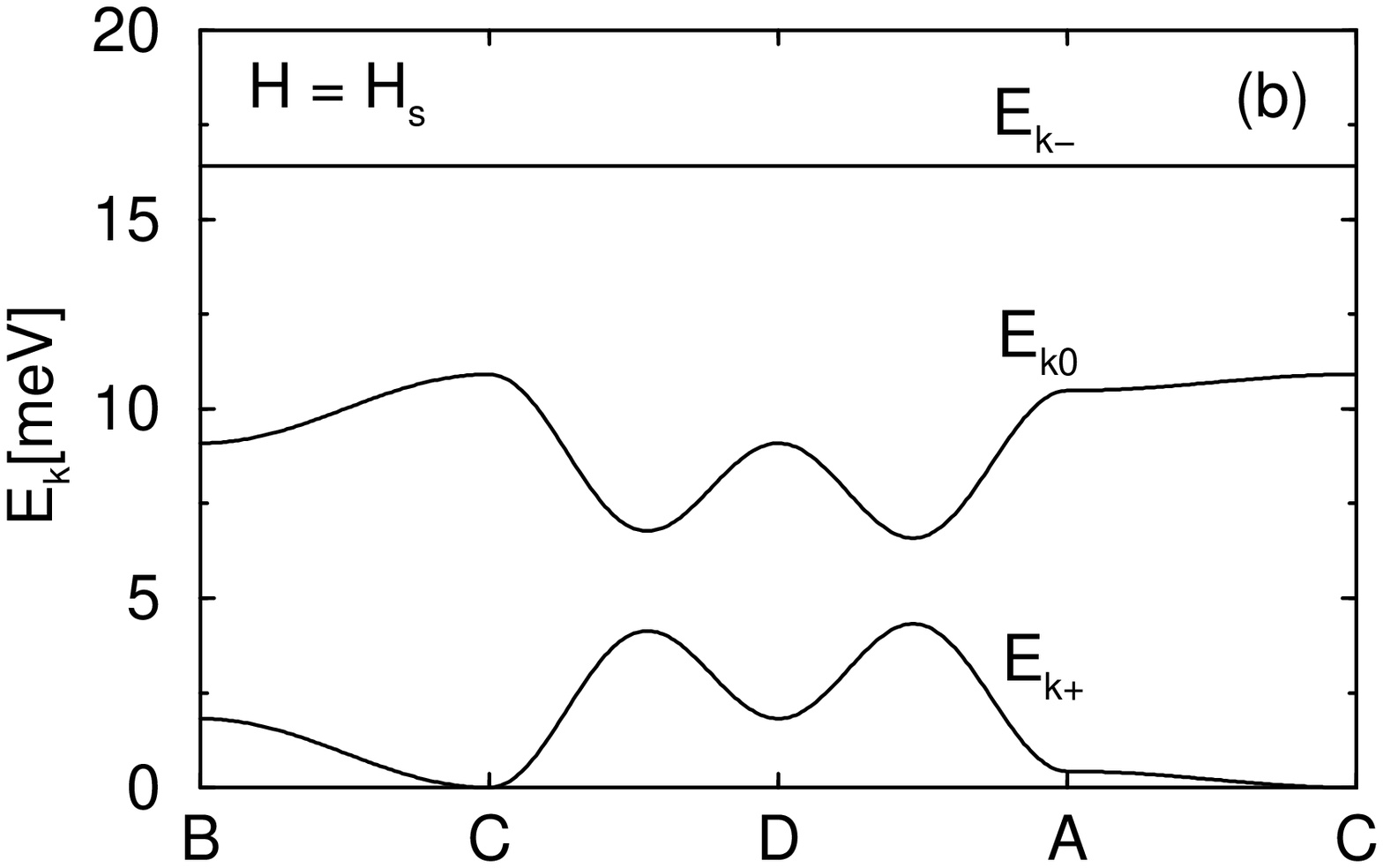}
\end{center}
\vspace{-0.5cm}
\caption{ Dispersion of the three magnon modes in \Tl: (a) at $H = 14{\rm T}
> H_c$ ({\it cf.} Ref.~\protect{\onlinecite{rrcfgkmwhv}}); (b) at the
saturation field $H = H_s$.}
\label{fig:dispersion-order}
\end{figure}

The $O(b^2)$ terms in Eq.~(\ref{eob2}) are diagonalized by two separate 
Bogoliubov transformations which yield the energies of the collective modes 
of the condensate. Because the triplet $b_{k0}$ does not mix with the other 
two modes, ${\cal H}_0$ is diagonalized easily to give  
\begin{equation} 
E_{\bsk 0} = \sqrt{\epsilon_{\bsk 0}^2 - \Delta_{\bsk 0}^2}, 
\label{eqn:dispersion-E2} 
\end{equation} 
which describes the middle magnon branch in the ordered phase. 
${\cal H}_{\pm}$ is diagonalized by the transformation
\begin{equation} 
\alpha_{\bsk}^m = u_{\bsk +}^m b_{\bsk +} + u_{\bsk -}^m b_{\bsk -} 
  + v_{\bsk +}^m b_{-\bsk +}^{\dag} + v_{\bsk -}^m b_{-\bsk -}^{\dag},
\label{epmdt} 
\end{equation} 
where $m = 1,2$ denotes the two modes corresponding to each value of $\bk$,
whence 
\begin{equation} 
{\cal H}_{\pm} = \sum_{\bsk m} E_{\bsk}^m \alpha_{\bsk}^{m \dag}  
\alpha_{\bsk}^m . 
\label{ehpmdt} 
\end{equation} 
$E_{\bsk}^m$ and $\alpha_{\bsk}^m$ are then the eigenvalue and eigenvector  
for the mode of branch $m$, and are related by the equation  
\begin{equation}
\left( \begin{matrix}
    \epsilon_{\bsk +} & -\Delta_{\bsk +} & \epsilon_{\bsk \pm} &
-\Delta_{\bsk \pm} \cr
    \Delta_{\bsk +} & -\epsilon_{\bsk +} & \Delta_{\bsk \pm} &
-\epsilon_{\bsk \pm} \cr
    \epsilon_{\bsk \pm} & -\Delta_{\bsk \pm} & \epsilon_{\bsk -} &
-\Delta_{\bsk -} \cr
    \Delta_{\bsk \pm} & -\epsilon_{\bsk \pm} & \Delta_{\bsk -} &
-\epsilon_{\bsk -} \cr \end{matrix} \right)
\left( \begin{matrix} u_{\bsk +}^m \cr v_{\bsk +}^m \cr
u_{\bsk -}^m \cr v_{\bsk -}^m \cr \end{matrix} \right) =
E_{\bsk}^m \left( \begin{matrix} u_{\bsk +}^m \cr v_{\bsk +}^m \cr
u_{\bsk -}^m \cr v_{\bsk -}^m \cr \end{matrix} \right).
\label{eqn:eigen}
\end{equation}
This equation has four eigenvalues, of which two are positive, 
and we define the smaller of these as $E_{\bsk}^+$, the dispersion  
relation of the lowest magnon mode, while the larger is defined as  
$E_{\bsk}^-$ for the highest mode. By the nature of the problem, the  
other two eigenvalues are $-E_{\bsk}^+$ and $-E_{\bsk}^-$. In the  
disordered phase, $\epsilon_{\bsk \pm} = 0 = \Delta_{\bsk +} =  
\Delta_{\bsk -}$, and the equation (\ref{eqn:eigen}) reduces to a  
$2\times 2$ matrix form.  

\begin{figure}[t!]
\begin{center}
\includegraphics[width=7cm]{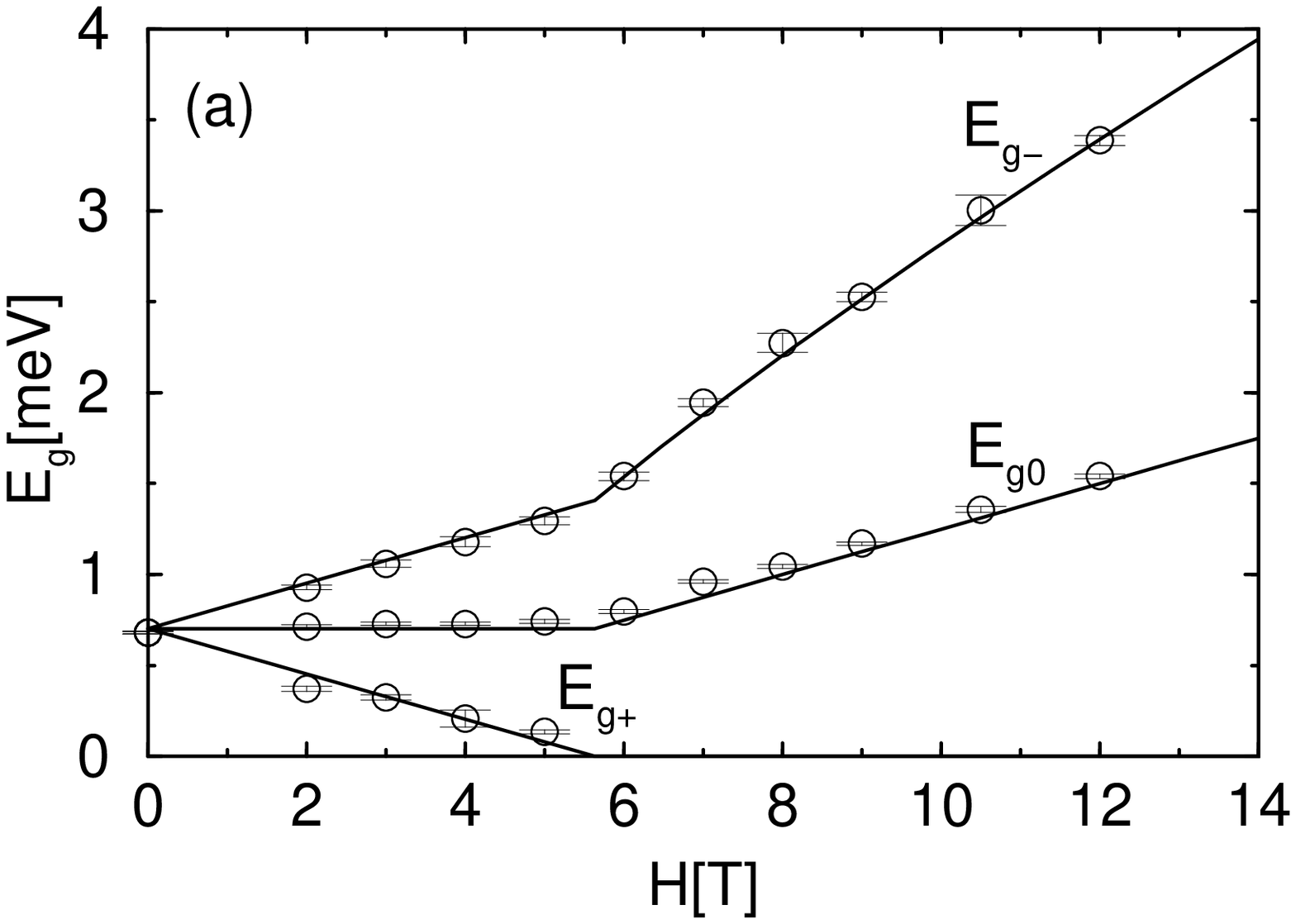}
\includegraphics[width=7cm]{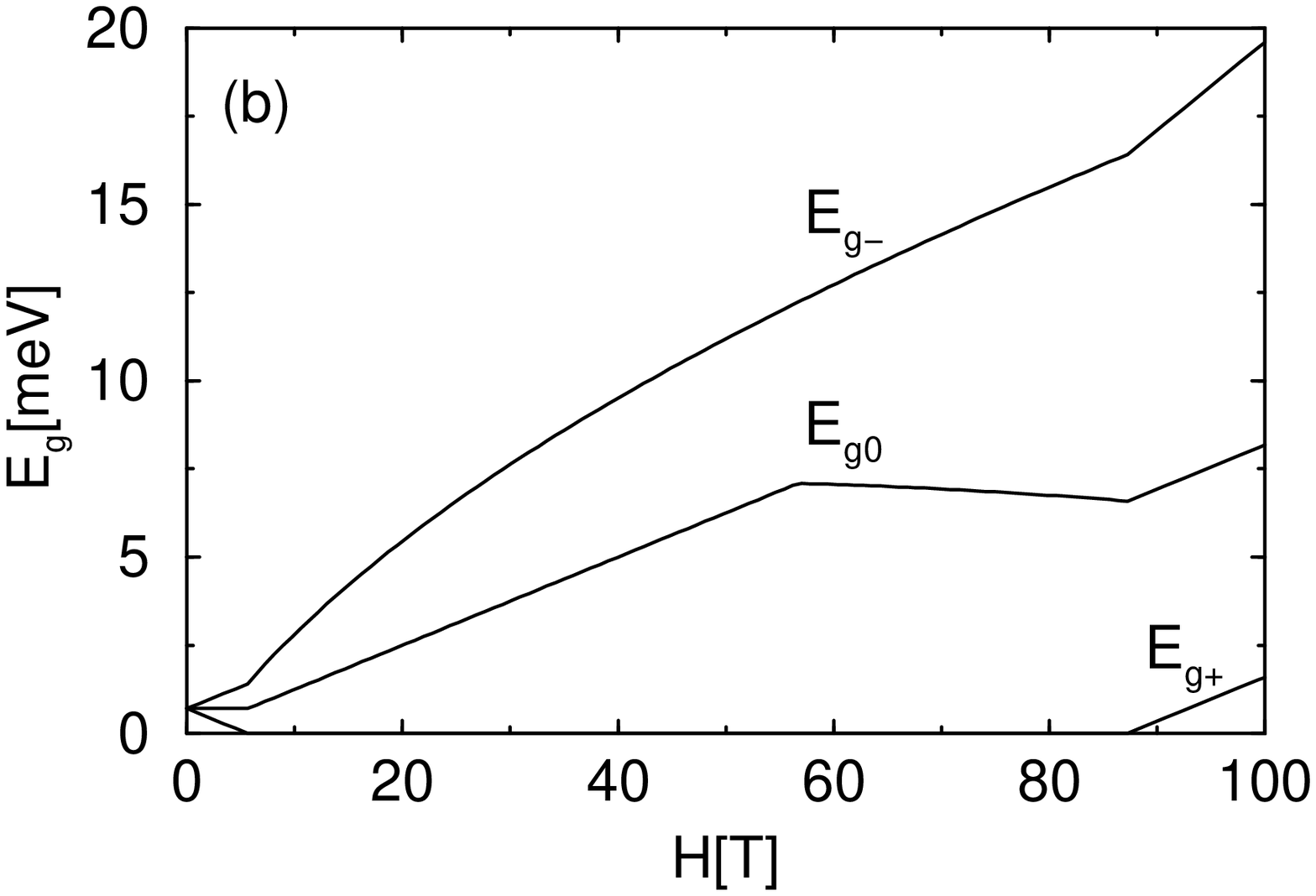}
\end{center}
\vspace{-0.8cm}
\caption{Field dependence of the energy gaps for the three magnon modes 
in \Tl: (a) points are data from the INS experiment of
Ref.~\protect{\onlinecite{rrcfkgvm}}, solid lines are the theoretical
fit using the zero--field parameters of Table~I; (b) field--dependence
at high fields, using $g = 2.16$ as the $g$--factor of \Tl.
\protect{\cite{rrcfkgvm}}}
\label{fig:gap-h}
\end{figure}

In the ordered phase above $H_c$, the mode $E_{\bsk +}$ is always massless  
at the band minimum, $\bQ$, and develops a linear dependence on $\bk$ (in  
all three directions in reciprocal space) in the vicinity of this point 
[Fig.~\ref{fig:dispersion-order}(a)]. This branch is the Goldstone mode,  
which arises from the fact that rotation of the staggered moment about  
the axis of broken symmetry (the field axis) does not change the energy of  
the system. It is the staggered magnetic order induced perpendicular to  
the applied magnetic field which breaks rotational symmetry around the  
field axis. Mathematically, rotations of the induced moment are realized 
by changing the phase of $f$ and $g$ in (\ref{eqn:b}) 
($f \rightarrow e^{-i\chi}f$, $g \rightarrow e^{i\chi}g$, where $\chi$  
is the rotation angle), and the invariance of the expressions under this  
transformation determines the gapless nature of the Goldstone mode. The  
remaining modes in Fig.~\ref{fig:dispersion-order}(a) show that for  
fields close to $H_c$ the overall shape of the dispersion relations is  
not strongly affected by the presence of the field while the field remains  
small on the scale of the mode energies $E_{\bsk}$.  
 
The field--dependence of magnon dispersion relations may be characterized  
up to high fields using the expressions for obtained for $E_{\bsk +}$,  
$E_{\bsk 0}$, and $E_{\bsk -}$ with the zero--field parameters of Table~I. 
Only in the disordered phase ($H \le H_c$) does the magnon dispersion not  
change its shape, with a simple linear splitting of the three branches  
developing due the Zeeman interaction until the energy gap of the lowest  
mode is reduced to zero at $H = H_c$. Above the critical field, the gap  
for $E_{\bsk +}$, the Goldstone mode of the ordered phase, remains
zero, as a consequence of which the gaps of the higher modes,  
$E_{g0}$ and $E_{g-}$, show an abrupt increase in slope at $H_c$  
[Fig.~\ref{fig:gap-h}(a), {\it cf.}~Fig.~\ref{fig:summary-field}.]. The 
results of the bond--operator theory (solid lines) show quantitative 
agreement with the experimental measurements obtained by INS.\cite{rrcfkgvm}

Fig.~\ref{fig:gap-h}(b) shows the predicted field dependence of the gaps  
for high fields: while $E_{g-}$ has monotonically increasing behavior  
all the way to and beyond $H_s$, the gap $E_{g0}$ of the middle branch  
first increases with field, but then decreases until $H_s$. This behavior  
is related to the position in the Brillouin zone of dispersion minimum,  
which is determined largely by the $\epsilon_{\bsk 0}$ term in  
the Appendix. As the magnetic field is raised, the sign of the dispersive 
component $\epsilon_{\bsk}$ changes on passing through the field at which 
$u^2 = v^2$, which for the parameters of \Tl~corresponds to 55T 
[Fig.~\ref{fig:u}(b)]. Below this value, where $u^2 > v^2$, the minimum  
is at $\bQ$, while for fields close to 55T ($u^2 \simeq v^2$) the dispersion  
becomes flat. Above this field ($u^2 < v^2$) the minimum moves from $\bQ$ to  
a point in the middle of the zone, as shown for the branch $E_{\bsk 0}$ in  
Fig.~\ref{fig:dispersion-order}(b), and $E_{g0}$ then decreases with field  
until saturation. Finally, at $H = H_s$, the $\bk$ dependence of the lowest  
mode, $E_{\bsk +}$, becomes quadratic again in the vicinity of $\bQ$  
[Fig.~\ref{fig:dispersion-order}(b)] as a gap opens to the lowest  
mode of the fully saturated ferromagnet [Fig.~\ref{fig:gap-h}(b)]. For 
$H > H_s$ (see below), the magnon modes decouple, the condensate has pure 
triplet character and the lowest excitation mode is a pure singlet whose  
gap grows linearly with the field, as do $E_{g0}$ and (with a slope greater  
by a factor of two) $E_{g-}$. The evolution with field of the shape of the  
dispersion relations through the ordered phase is illustrated by  
Fig.~\ref{fig:dispersion-order}, where it is clear that the lower branch  
changes rather little, the upper branch becomes progressively narrower  
until it is completely non--dispersive at $H_s$, and the middle branch  
is inverted as discussed above.  

\begin{figure}[t!]
\begin{center}
\includegraphics[width=3.7cm]{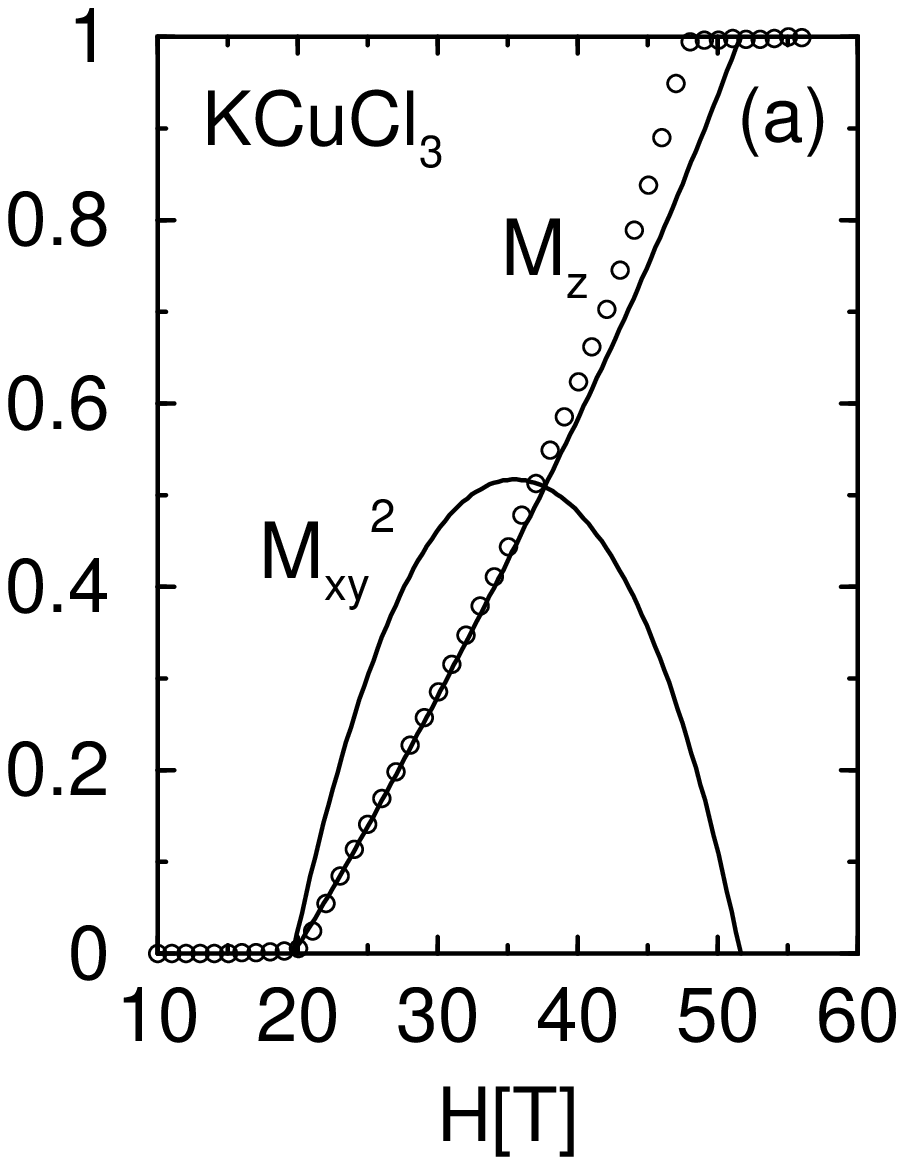}
\includegraphics[width=3.8cm]{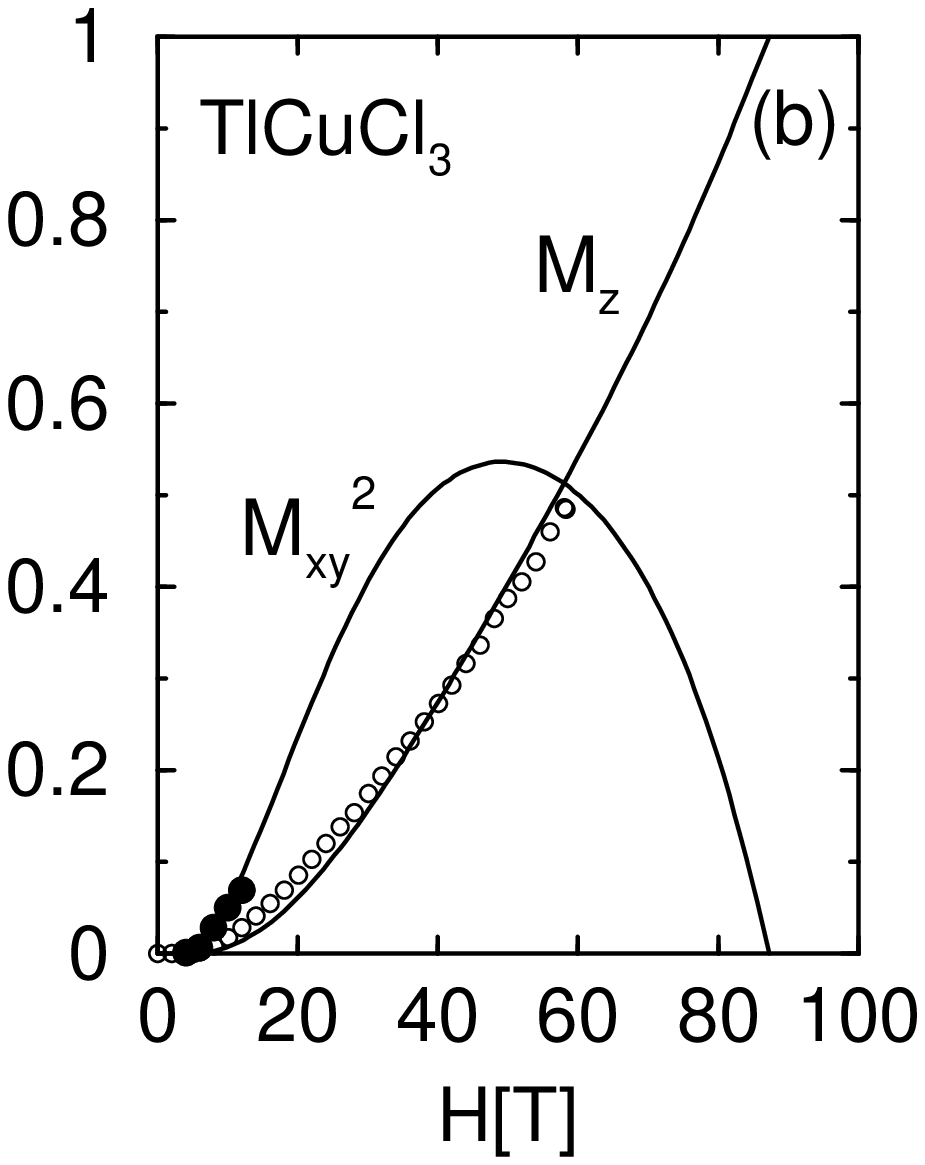}
\end{center}
\vspace{-0.8cm}
\caption{(a) Normalized magnetization curves for \K.
Open circles are experimental data for $M_z$ measured at $T =
1.3$K.\protect{\cite{rotatstktk}} The theoretical fits are made using
$g = 2.29$ as the $g$--factor for \K.\cite{rotatstktk}
(b) Normalized magnetization curves for \Tl. Open circles are
experimental data for $M_z$ measured at $T = 1.3$K,\protect{\cite{rtokuokh}}
and closed circles are data for $M_{xy}$ measured at $T =
0.2$K.\protect{\cite{rtkot}} The theoretical fits are made using the
same $g$--factor for \Tl~as in Fig.~\protect{\ref{fig:gap-h}}.
}
\label{fig:M}
\end{figure}

\subsubsection*{Magnetization}

The theoretical framework also gives direct access to the average
magnetization of the dimers in the ordered--phase condensate. We denote by  
$M_z$ the magnetization parallel to the field, whose normalized value is  
\begin{equation} 
M_z = \frac{1}{N} \sum_i \langle t_{i +}^\dag t_{i +} 
                                -t_{i -}^\dag t_{i -} \rangle,  
\label{emz} 
\end{equation} 
and by $M_{xy}$ the perpendicular (staggered) magnetization
\begin{eqnarray} 
M_{xy} & = & \frac{1}{N} \sum_i e^{i\bsQ\cdot\br_i} \langle S_{ilx} 
- S_{irx} \rangle \label{emxy} \nonumber \\ & = & \frac{1}{N} \sum_i 
e^{i\bsQ\cdot\br_i} \langle s_i^\dag (t_{i+} + t_{i-}) +{\rm H.c.} \rangle, 
\end{eqnarray}
where $i$ labels the dimers and $\langle \dots \rangle$ signifies
an expectation value calculated using the operators $\alpha_{\bsk}^m$ in 
Eq.~(\ref{epmdt}). The mean--field magnetizations, normalized to their 
values at saturation are given by 
\begin{eqnarray} 
M_z & = & v^2 (f^2 - g^2), \label{emfml} \\ 
M_{xy} & = & {\textstyle \sqrt{2}} u v (f + g). \label{emfmt} 
\end{eqnarray} 
These expressions correspond to ${\overline a}^2$ contributions, and are 
supplemented by terms of the form $\alpha_\bsk \alpha_\bsk^\dagger$, 
which correspond to quantum corrections due to magnon excitations. Both 
contributions are shown in Figs.~\ref{fig:M}(a) and \ref{fig:M}(b), but 
the quantum corrections are found to be small, presumably as a result of 
the three--dimensionality of the \K~and \Tl~systems. In both cases the 
square of the staggered moment ($M_{xy}^2$) shows a linear field dependence 
close to $H_c$ and to $H_s$, indicating that $M_{xy} \sim \sqrt{H - H_c}$ as 
expected from a mean--field description.

For \K,~the magnetization parallel to the field ($M_z$) is almost 
perfectly linear in $H$ [Fig.~\ref{fig:M}(a)], in good agreement with 
experiment.\cite{rotatstktk} If one neglects the upper triplet mode ($t_-$) 
in the condensate specified by Eq.~(\ref{eqn:b}), $M_z$ would 
be completely linear in $H$, and because all interdimer interactions are 
small (Table I), \K~is close to this limit. By contrast, the interladder 
interaction $J_{2}$ is considerably larger for \Tl, where mixing into the 
condensate of the $t_-$ mode, mediated primarily by interactions of the type 
$t_+^\dag t_-^\dag s s$, is significant. This point can be seen clearly by 
comparing Figs.~\ref{fig:u}(a) and \ref{fig:u}(b), where the value of $g^2$ is 
considerably larger for \Tl~than for \K~in the vicinity of $H_c$. Physically,
the presence of the highest triplet in the condensate costs Zeeman energy, 
which leads to a suppression of $v$, and consequently $M_z$ is reduced near  
$H_c$ for \Tl. This causes a significant deviation (due in fact to higher  
powers in $H - H_c$) from purely linear field dependence, as illustrated in  
Fig.~\ref{fig:M}(b), which is again in very good agreement with the  
observed form.\cite{rsttktmg} This result emphasizes the importance of  
including the $t_-$ triplet in a consistent description of the ground  
state, which is required to account for the difference between the  
magnetization curves of \Tl~and \K. 
 
\subsection{Saturated phase ($H \ge H_s$)} 
 
In the high--field regime the spins are fully polarized. In the  
notation of Sec.~IIIB, the saturated phase is described by $(\theta,\phi)  
= (\pi/2,0)$ in the Hamiltonian of Eq.~(\ref{eqn:hamiltonian-order}),  
which specifies that the ground state becomes a pure condensate of the  
lowest triplet ($t_+$ in the notation of Sec.~IIIA). 
This phase is characterized by a complete decoupling of the triplet modes,  
by a finite energy gap to the lowest mode, which corresponds to an excitation  
from the triplet ground state to a singlet, and by the fact that the highest  
triplet mode becomes dispersionless [Fig.~\ref{fig:dispersion-order}(b)]. 
In this regime the interaction term ${\cal H}_{\rm tt}$ between the lowest  
and highest triplet branches in Eq.~(\ref{eqn:Htt}) becomes an
on--site potential term for $t_-$ magnons, and does not assist their  
propagation, so there is no first--order hopping process for this branch.  
A possible second--order process exists for propagation of the upper  
triplet by creation of two singlets in a virtual process due to the  
term $s_i^\dag s_j^\dag t_{j+} t_{i-}$ in ${\cal H}_{\rm st}$. The  
quantitative effect of this term on the dispersion $E_{\bsk -}$ is  
expected to be very small, but the process may also have a qualitative  
role in mediating the decay of $t_-$ magnons into two singlet modes (bearing  
in mind the triplet nature of the condensate). With energy and momentum  
conservation specified according to  
\begin{equation} 
E_{\bsk -} = \sum_\bsq (E_{\bsk+\bsq,+} + E_{\bsk-\bsq,+}), 
\end{equation} 
this may lead to a finite lifetime for the highest triplet excitations. 
We note also that this process is possible for fields $H_c \le H \le H_s$, 
and provides one reason for which the upper magnon mode may become broad 
at high magnetic fields. 
 
\section{Pressure--induced order} 
 
In Sec.~III we have discussed field--induced magnetic order in a  
quantum spin system, which can be understood qualitatively as the  
consequence of the lowering of one of the triplet modes by the applied  
magnetic field to the point where it becomes soft, and the ordered  
phases is favored. Another way in which the magnon modes may soften,  
causing the onset of magnetic order, is by increasing the interdimer 
interactions, broadening the magnon band. The application of pressure 
is expected to have just this effect, and for the small--gap system  
\Tl~this type of pressure--induced AF order was found recently by  
Oosawa {\it et al.}\cite{rofokt} in neutron diffraction experiments  
conducted under a fixed hydrostatic pressure $P = 1.48{\rm GPa}$. 
 
As noted in Sec.~III, this type of ``pressure--induced'' magnetic QPT  
provided the first application of the bond--operator formulation to ordered  
phases.\cite{rnr2} In the disordered phase the three magnon modes are
degenerate at $H = 0$ for all couplings, including at the QCP where they 
are gapless at the wave vector $\bQ$ characterizing the incipient magnetic 
order. The excitations of the ordered AF consist of two spin waves, which are  
the conventional phase modes (Goldstone modes) of the ordered moment, and  
a third which corresponds to its amplitude fluctuations and is generally  
not relevant for the low--energy dynamics, but may become low--lying  
close to the QCP.\cite{rnr2} The ordered phase is described in the  
bond--operator approach by a condensation of one of the triplet operators,  
conventionally taken as $t_z^{\dag}$ in Eq.~(\ref{ebor}), and in  
Ref.~\onlinecite{rnr2} this was treated as a separate condensate in  
order to obtain an effective description of the ordered phase close to  
the QCP.
 
A complete account of the changes to the superexchange interactions caused 
by the application of a hydrostatic pressure would require a detailed  
structural investigation of the alterations to bond lengths and angles. 
Such a study lies beyond the scope of the current analysis, the aim of  
which is to characterize the changes in physical properties through the  
QPT. We thus employ qualitative arguments based on the considerations  
of Sec.~II for the probable evolution of the coupling constants in the  
XCuCl$_3$ structure, and use this as the foundation for a plausible model.  

The structural unit composed of CuCl$_6$ octahedra is not expected to 
undergo significant changes except at rather high pressures, and so we 
assume here that the dimer coupling $J$ is unaltered. The leading effects 
of hydrostatic pressure are expected to be a small longitudinal compression 
of the ladder units, and, due to the absence of strong covalent bonds in 
the transverse directions, a somewhat larger reduction in their separation.
These changes generally involve small reductions in bond lengths and more 
significant distortions of the bond angles. Considering next the effective 
interdimer coupling ${\tilde J}_1 = 2 J_1 - J_1^{\prime}$ (in the notation 
of Sec.~II), any small increases in ladder 
leg coupling $J_1$ are likely to be offset to a considerable extent by 
similar changes in the diagonal coupling $J_1^{\prime}$, so we will also
neglect the pressure--induced alteration of this parameter in the following 
simplified treatment. The leading pressure effects are thus expected to 
involve the alteration of the parameters $J_2^{\prime}$ and $J_3$, both of 
which are determined by Cu--Cl--Cl--Cu pathways, although the latter may 
also have contributions from the X$^+$ ion. We assume (based on the 
experimental observation of the QPT) that $J_2$ and $J_3$ increase with the 
pressure, $P$. For the functional form of this increase, the exact dependence 
of superexchange interactions on bond lengths and angles remains a topic 
active research, and in the absence of structural data under pressure we 
will attempt to fit this form with minimal assumptions. We take the 
interladder interactions to change with applied pressure according to 
\begin{eqnarray}
J_2(P) & = & p(P) J_2, \label{ej23p} \nonumber \\
J_3(P) & = & p(P) J_3,
\end{eqnarray}
where $p(P)$ is a dimensionless function of the real pressure $P$ and the 
values of $J_2$ ($J_2''$) and $J_3$ (${\tilde J}_3$) at $p = 1$ are listed 
in Table I. We begin by analyzing the situation where the interladder 
interactions are simply increased by a factor $p$, and then discuss the 
experimental determination of the function $p(P)$.

\subsection{No magnetic field}

\begin{figure}[t!]
\begin{center}
\includegraphics[width=7cm]{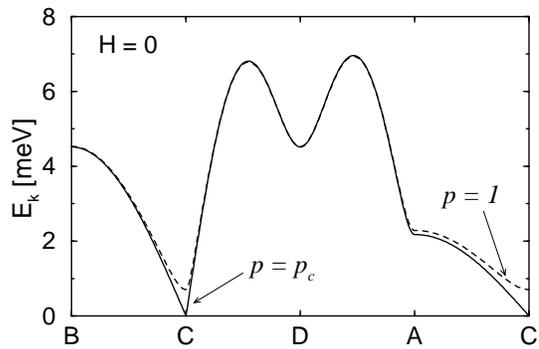}
\end{center}
\vspace{-0.5cm}
\caption{Threefold degenerate triplet magnon dispersion at zero field.
Dashed and solid lines correspond respectively to atmospheric ($p = 1$)
and critical ($p = p_c\simeq 1.018$) pressures.}
\label{fig:dispersion-pressure-disorder}
\end{figure}

With a view to the simultaneous application of field and pressure (below),  
we discuss the pressure--induced QPT using the triplet operators $t_+$, 
$t_0$, and $t_-$ (\ref{eqn:transformation-1}) which were introduced for 
the field--induced QPT, rather than in the ``natural'' basis of operators 
$t_x$, $t_y$, and $t_z$ (\ref{ebor}). In the absence of a magnetic field, 
the magnon dispersion in the disordered phase is given  
by Eq.~(\ref{eqn:dispersion}) with $h = 0$, and the single branch is  
triply degenerate. The dispersion, shown at atmospheric pressure by the  
dashed line in Fig.~\ref{fig:dispersion-pressure-disorder} ({\it cf.}  
Fig.~\ref{fdt}), is given by  
\begin{equation} 
E_\bsk = \sqrt{J^2 + 2J\epsilon_\bsk}, 
\label{eqn:dispersion-2} 
\end{equation} 
where $\epsilon_\bsk$ is defined in (\ref{eqn:dispersion0}) but with $J_2$  
and $J_3$ now given by Eq.~(\ref{ej23p}). As pressure is increased, the  
maximal positive and negative values of $\epsilon_\bsk$ are enhanced, the  
bandwidth increases and the dispersion minimum (spin gap) decreases. It is  
clear from Fig.~\ref{fig:dispersion-pressure-disorder} that the low--energy  
part of the magnon spectrum is particularly sensitive to the pressure,  
whereas the higher--energy parts are scarcely affected. This result is a  
simple consequence of the square--root form of Eq.~({\ref{eqn:dispersion-2}),
which amplifies the changes in $E_\bsk$ where the two terms are close to  
mutual cancellation, {\it i.e.}~in the vicinity of the band minimum $\bQ$. 
 
The magnon modes become soft, $E_\bsQ = 0$, at a critical pressure, $p_c$,  
where the enhanced interladder interactions favor a long--ranged AF spin 
order over a gapped dimer phase. The dimensionless pressure $p(P)$ at the 
critical pressure for the parameters of \Tl~is  
\begin{equation} 
p_c = \frac{J - J_1}{J_2 + 2J_3} \simeq 1.018, 
\end{equation} 
from which one notes immediately that the critical pressure represents only 
a very small fractional increase in $p$. 

Expansion of $\epsilon_\bsk$ in Eq.~(\ref{eqn:dispersion-2}) around $\bQ$, 
$\epsilon_{\bsQ + \delta\bsk} \simeq \epsilon_\bsQ + \delta \epsilon_\bsk$, 
yields a dispersion  
\begin{equation} 
E_{\bsQ + \delta\bsk} \simeq \sqrt{E_\bsQ^2 + 2J\delta\epsilon_\bsk}. 
\end{equation} 
Because $\epsilon_\bsk$ contains cosine terms in $\bk$ which are  
minimal at $\bQ$, $\delta \epsilon_\bsk$ is quadratic in $\delta\bk$,
and at the critical pressure $E_\bsk$ has a linear $\bk$--dependence 
around $\bQ$ [Fig.~\ref{fig:dispersion-pressure-disorder}].
This stands in explicit contrast to the case of field--induced order,  
where the single low--lying magnon retains a quadratic dependence on  
$\bk$ at the critical field where it softens. 
 
\begin{figure}[t!] 
\begin{center} 
\includegraphics[width=7cm]{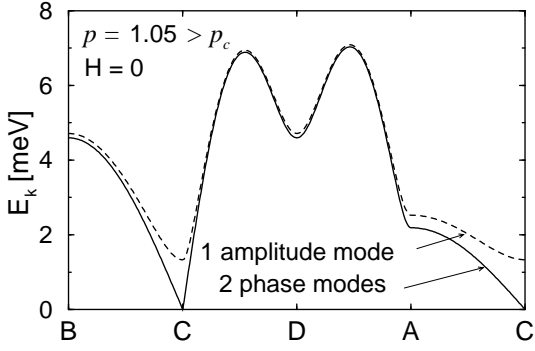} 
\end{center} 
\vspace{-0.5cm}
\caption{ Magnon dispersion relations in pressure--induced AF phase 
at zero field, showing two degenerate phase modes (solid line) and one 
amplitude mode (dashed line). } 
\label{fig:dispersion-amplitude} 
\end{figure} 
 
Increasing the pressure beyond $p_c$ induces the AF ordered phase,  
which may be described by the same operator transformation as in Sec.~III. 
In the absence of an applied magnetic field, the coupled triplets $t_+$
and $t_-$ are equivalent and $f = g = 1/\sqrt{2}$ ($\phi = \pi/4$) 
in Eq.~(\ref{eqn:b}). We note here that this combination in  
fact restores the eigenvectors $t_x$ and $t_y$ in Eq. (\ref{ebor}), 
referred to  above as the ``natural basis'' for the field--free problem. 
Using the $b$ operators defined in Eq.~(\ref{eqn:b}), the Hamiltonian at 
quadratic order takes the symmetrical form  
\begin{equation} 
{\cal H}_0 + {\cal H}_{\pm} = \sum_{\alpha=+,0,-} {\cal H}_\alpha 
\label{ehqpop} 
\end{equation} 
with  
\begin{equation} 
{\cal H}_\alpha = \sum_\bsk [ \epsilon_{\bsk \alpha} b_{\bsk  
\alpha}^\dag b_{\bsk \alpha} + {\textstyle \frac{1}{2}} (\Delta_{\bsk  
\alpha}^\dag b_{\bsk \alpha} b_{-\bsk \alpha}+{\rm H.c.})].   
\label{ehqpopc} 
\end{equation}
The matrix elements are given by  
\begin{eqnarray} 
\epsilon_{\bsk +} & = & J(u^2 - v^2) - 8\epsilon_\bsQ u^2 v^2 +(u^2 - v^2)^2  
\epsilon_\bsk, \label{ehqpopcc} \nonumber \\ 
\epsilon_{\bsk 0} & = & \epsilon_{\bsk -} \,\, = \,\, J u^2 - 4 \epsilon_\bsQ  
u^2 v^2 + (u^2 - v^2) \epsilon_\bsk, \nonumber \\ 
\Delta_{\bsk +} & = & (u^2 - v^2)^2 \epsilon_\bsk, \\ 
\Delta_{\bsk 0} & = & -\Delta_{\bsk -} \,\, = \,\, \epsilon_\bsk, \nonumber  
\end{eqnarray} 
where  
\begin{equation} 
u^2 = \frac{1}{2} + \frac{J}{(-4\epsilon_\bsQ)}, \;\;\; v^2 = \frac{1}{2}  
- \frac{J}{(-4\epsilon_\bsQ)}  
\label{ehqpopuv} 
\end{equation} 
and  
\begin{equation} 
\epsilon_\bsQ = - {\textstyle \frac{1}{2}} \left[J_1 + p(J_2 + 2 J_3)\right].
\label{ehqpopeq}
\end{equation}
The magnon mode dispersion relations are then given by 
\begin{equation} 
E_{\bsk\alpha} = \sqrt{\epsilon_{\bsk\alpha}^2 - \Delta_{\bsk\alpha}^2}, 
\label{eqn:dispersion-pressure} 
\end{equation} 
and are illustrated in Fig.~\ref{fig:dispersion-amplitude}. With this 
choice of coordinates the staggered magnetic moment is developed along
the ${\hat x}$--axis in spin space, as the transformed magnon operator  
$b_{\bsk+}$ is precisely $t_x$ of Eq.~(\ref{ebor}). It is clear that  
the other two modes remain degenerate, and that $E_{\bsk 0}$ and  
$E_{\bsk -}$ are gapless at $\bQ$ with linear dispersion around this  
point. These are the spin waves of the ordered magnet, or Goldstone modes; 
their magnetic moment fluctuates in the plane perpendicular to the induced  
moment, demonstrating their role as the transverse phase modes. The remaining  
mode, $E_{\bsk +}$, has a finite gap and corresponds to fluctuations in the  
direction of the induced moment, confirming that this is the longitudinal,  
or amplitude, mode of the pressure--induced AF phase. This is the  
splitting of the threefold degenerate triplet modes expected from  
previous studies of coupling--induced AF order in gapped spin  
systems.\cite{ra,raw,rnr2} The excitation gap of the amplitude mode may be  
written as 
\begin{eqnarray} 
E_{\bsQ +} & = & 2 u v \sqrt{2 + u^2 v^2} \label{esgam} \\ & = & 2  
\sqrt{\left[\frac{1}{4} - \left(\frac{J}{4\epsilon_\bsQ}\right)^2 \right] 
\left[2 + \frac{1}{4} - \left(\frac{J}{4\epsilon_\bsQ}\right)^2 \right]},  
\nonumber 
\end{eqnarray} 
and its evolution with applied pressure is shown in  
Fig.~\ref{fig:gap-amplitude}. 
 
\begin{figure}[t!] 
\begin{center} 
\includegraphics[width=7cm]{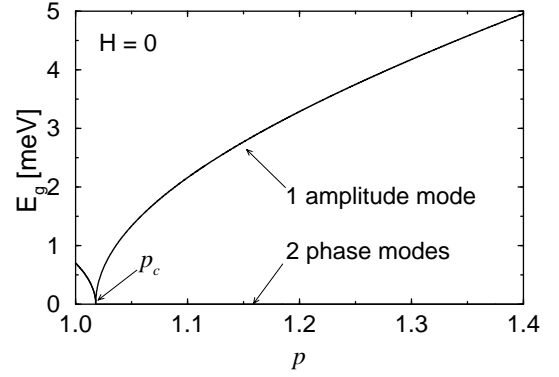} 
\end{center} 
\vspace{-0.8cm}
\caption{ Pressure dependence of spin gaps for phase and amplitude modes.} 
\label{fig:gap-amplitude} 
\end{figure} 

\begin{figure}[t!]
\begin{center}
\includegraphics[width=7cm]{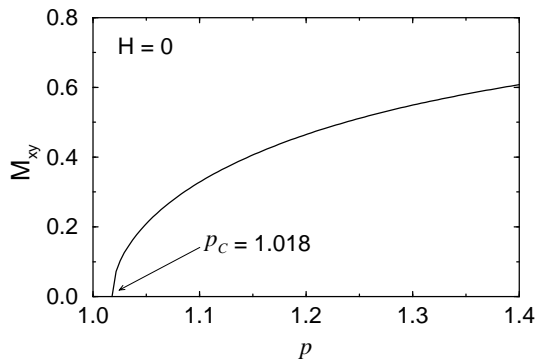}
\end{center}
\vspace{-0.8cm}
\caption{ Pressure dependence of induced staggered moment, $M_{xy}$,
in \Tl~at zero field. The normalization is to a saturated moment of 1.}
\label{fig:M-pressure}
\end{figure}

We turn next to a discussion of the pressure dependence of the interladder 
superexchange parameters, {\it i.e.}~to a determination of $p(P)$. 
The staggered moment grows with increasing pressure in the ordered phase 
as shown in Fig.~\ref{fig:M-pressure}, and close to the critical pressure 
it has the mean--field exponent $M_{xy} \propto (p - p_c)^{1/2}$.
Unlike the field--induced order, there is no uniform magnetization at zero  
field, which is reflected in the theoretical description by the relation 
$f = g$. The critical pressure has been found to be approximately $P_c = 
2$kbar.\cite{rtpc} By expanding the function at the critical pressure, a 
crude determination of $p(P)$ may be made by assuming the form
\begin{equation}
p(P) - p_c = \beta_1 (P - P_c) + \beta_2 (P - P_c)^2,
\end{equation}
where $\beta_1$ and $\beta_2$ are constants to be determined from 
\begin{eqnarray}
p_0 - p_c &=& \beta_1 (P_0 - P_c) + \beta_2 (P_0 - P_c)^2, \\
p_1 - p_c &=& \beta_1 (P_1 - P_c) + \beta_2 (P_1 - P_c)^2.
\end{eqnarray}
The three known data points are $i)$ atmospheric pressure $P = P_0$, where 
$p_0 = 1$, $ii)$ the critical pressure $P = P_c = 2$kbar, where $p = p_c 
= 1.018$, and $iii)$ $P = P_1 = 1.48$GPa (14.8kbar), where the staggered 
magnetization $M_{xy}$ has attained  60\% of its saturation value,\cite{rofokt}
which corresponds to $p_1 = 1.4$ (Fig.~\ref{fig:M-pressure}). With these 
values the prefactors of the lowest orders in the expansion are $\beta_1 
= 0.012 $kbar$^{-1}$ and $\beta_2 = 0.0014$kbar$^{-2}$. The relative 
decrease in magnitude of these coefficients suggests that higher orders may 
not be significant, although a detailed experimental determination is 
awaited. In the vicinity of the critical pressure, the finite linear term 
in the pressure dependence of $p(P)$ leads to an exponent $M_{xy} \propto 
(P - P_c)^{1/2}$. The most important qualitative observation arising from 
these considerations is that obtaining a critical pressure of $P_c = 2$kbar 
is straightforward in the laboratory, which in practice implies that 
experiments probing the properties of the \Tl~system on both sides of the 
pressure--induced QPT could be performed on samples of considerable volume.

The pressure dependence of the spin gap of the amplitude mode (\ref{esgam}),  
shown in Fig.~\ref{fig:gap-amplitude} is similar to that of the induced  
staggered moment (Fig.~\ref{fig:M-pressure}), as both depend in a similar  
manner on the singlet--triplet mixing coefficient $\theta$. Under a pressure
of 1.48GPa, where the neutron diffraction experiment revealing magnetic  
order was performed, we estimate the gap of the amplitude mode to be 5meV.  
This is well above the value required to distinguish the amplitude mode  
from the spin waves and magnetic Bragg peaks in INS experiments, which may 
thus be performed at lower pressures and with larger samples. \Tl~therefore 
offers an ideal system for observation of the amplitude mode, and for the 
measurement of its dispersion over a significant portion of the Brillouin 
zone. These measurements could be used to test the hypothesis that the 
amplitude mode is broad, and potentially ill--defined, due to the 
possibility of decay processes into pairs of spin waves.\cite{rmzpc} The 
gap of the amplitude mode may also be measured from the location of 
additional contributions to susceptibility, nuclear magnetic resonance 
(NMR), and electron spin resonance (ESR) signals.

We conclude this section by noting the two qualitative differences between 
the pressure--induced QPT and the field--induced situation discussed in 
Sec.~III. The first concerns the breaking of symmetry in real space and 
in spin space. In the field--induced case the plane of the staggered moment 
is dictated by the applied magnetic field, and its direction within the 
plane is determined by a spontaneous breaking of the $O(2)$ symmetry. 
Rotation of the moment corresponds to the single Goldstone mode. In the 
pressure--induced case there is no coupling of the spin direction to the 
crystal axes (in the absence of additional terms in the Hamiltonian which 
are not introduced here), the development of AF order requires a 
spontaneous breaking of $O(3)$ symmetry, and two Goldstone modes specify 
the rotations of the staggered moment. Here we have for convenience chosen 
the ${\hat x}$--direction, but the results are identical to bond--operator 
treatments choosing any other symmetry--breaking axis. The second 
difference concerns the triplet modes participating in the condensate 
in the bond--operator description: in the field--induced case the applied 
magnetic field lowers the symmetry of the system in a manner such that 
only one mode is driven to become massless at the QPT. It is this mode 
whose mixing into the singlet determines both the primary ground--state 
properties of the ordered phase and the nature of the phase fluctuation
(the Goldstone mode), while the small admixture of the highest triplet is 
required primarily for consistency. In the pressure--induced case the 
degeneracy of the magnons is lifted only by the spontaneous breaking 
of symmetry, and the condensate is composed only of the singlet and the 
triplet ($t_x$) which constitutes the amplitude mode, {\it i.e.}~the 
Goldstone modes (spin waves) explicitly do not form a part of the condensate. 
These two features are responsible for the differences between pressure-- and 
field--induced QPTs at the level of quantities such as critical exponents 
(below). Because the field acts to break the symmetries responsible for the 
properties of the pressure--induced QPT, any studies of quantum criticality 
under simultaneous application of finite field and pressure are dominated by 
the behavior of the field--induced QPT.

\begin{figure}[t!]
\begin{center} 
\includegraphics[width=7cm]{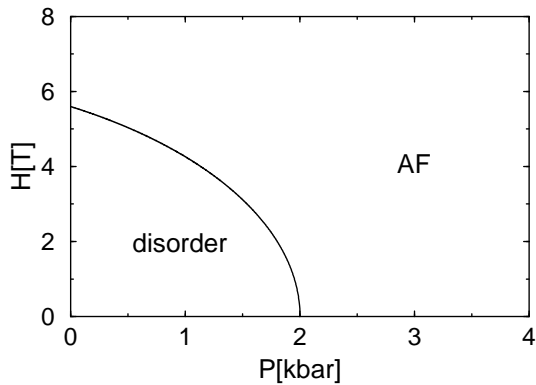}
\end{center} 
\vspace{-0.8cm}
\caption{Phase diagram for the model in the $P$--$H$ plane. } 
\label{fig:phase-diagram} 
\end{figure} 
 
\subsection{Finite magnetic field}

Magnetic fields and hydrostatic pressure may be regarded as cooperating  
factors in the stabilization of an AF ordered phase. One may thus  
determine the phase boundary between magnetically ordered and disordered  
phases, and this is shown in the estimated $P$--$H$ phase diagram in  
Fig. \ref{fig:phase-diagram}. The phase boundary is determined by the  
condition 
\begin{equation} 
g\muB H = \sqrt{J^2 - J [ J_1 + p(P)(J_2 + 2J_3) ]}.
\end{equation}
The leading linear dependence of the superexchange on pressure in the 
vicinity of $P_c$ results in the phase boundary having the form $H_c 
\propto (P_c - P)^{1/2}$. We note that the phases on both sides 
of the boundary show considerable evolution across the phase diagram in 
properties such as the uniform and staggered magnetic moments, and the 
splitting and dispersion of the magnon modes, as a function of the ratio 
of $H$ to $P$.

\begin{figure}[b!]
\begin{center}
\includegraphics[width=7cm]{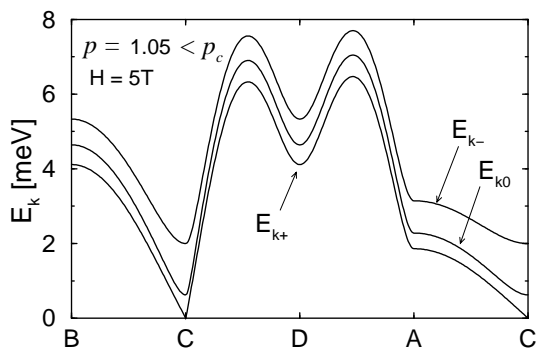}
\end{center}
\vspace{-0.5cm}
\caption{Magnon dispersion relations in the pressure--induced AF phase
under a magnetic field $H = 5{\rm T} < H_c$. }
\label{fig:dispersion-pressure-h}
\end{figure}

As noted in the preceding subsection, the degeneracy and dispersion of  
the magnon modes in the pressure--induced ordered phase are altered  
significantly in the presence of a magnetic field. A finite field lifts  
the degeneracy of the transverse spin waves because their spin fluctuation 
directions ($y$ and $z$ here) are no longer equivalent. The field causes 
the mixing of the amplitude mode, $E_{\bsk +}$, with the phase mode 
$E_{\bsk -}$, and the mixing coefficient  
changes from $\phi = \pi/4$ at zero field to $\phi = 0$ in the high--field  
limit, where the modes $t_+$ and $t_-$ are again the appropriate  
eigenstates. Fig.~\ref{fig:dispersion-pressure-h} shows the dispersion  
relations for the three nondegenerate modes at a pressure $p > p_c$ for
a finite magnetic field: the qualitative features of the spectra are 
manifestly dominated by the applied field, and only one Goldstone  
mode remains. In Fig.~\ref{fig:dispersion-pressure-h} we have redefined  
the lower and upper magnon branches as $E_{\bsk +}$ and $E_{\bsk -}$,  
respectively, as in the study of field--induced order (Sec.~III). 

\begin{figure}[t!]
\begin{center}
\includegraphics[width=7cm]{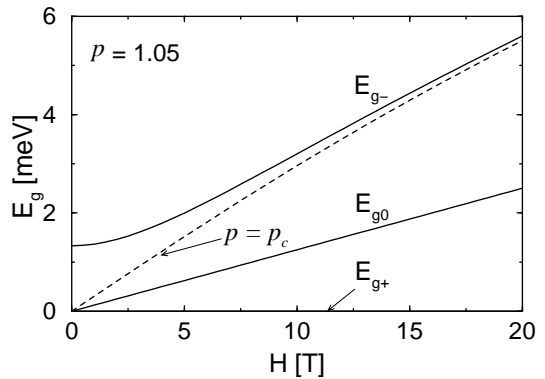}
\end{center}
\vspace{-0.8cm}
\caption{Magnetic--field dependence of spin gaps in the pressure--induced 
AF phase, illustrated for $p=1.05$. The dashed line shows the behavior of 
the highest mode at the critical pressure. The spin gap of the middle 
branch, which is a spin wave at $h = 0$, is the same for $p = p_c$ and
$p > p_c$. }
\label{fig:gap-pressure-h}
\end{figure}

Fig.~\ref{fig:gap-pressure-h} shows the field dependence of the spin gaps 
of the three branches in the pressure--induced ordered phase $p > p_c$,  
with the same labeling convention as before. The middle branch
does not mix with the upper and lower branches even at a finite field,
so $E_{g0}$ shows a linear increase from zero which is in fact independent  
of $p$ for all $p \ge p_c$. Because the amplitude mode has a finite  
excitation gap for $p > p_c$ at $h = 0$, the gap grows only quadratically  
at small fields (linearly at $p = p_c$). It is clear that the amplitude  
mode is adiabatically connected to the upper magnon branch of the  
field--only problem ($p < p_c$), which for $p > p_c$ is retrieved in the 
high--field limit by a continuous alteration of the mixing coefficient 
$\phi$ under the external magnetic field.

\subsection{Magnetization}

\begin{figure}[t!]
\begin{center}
\includegraphics[width=7cm]{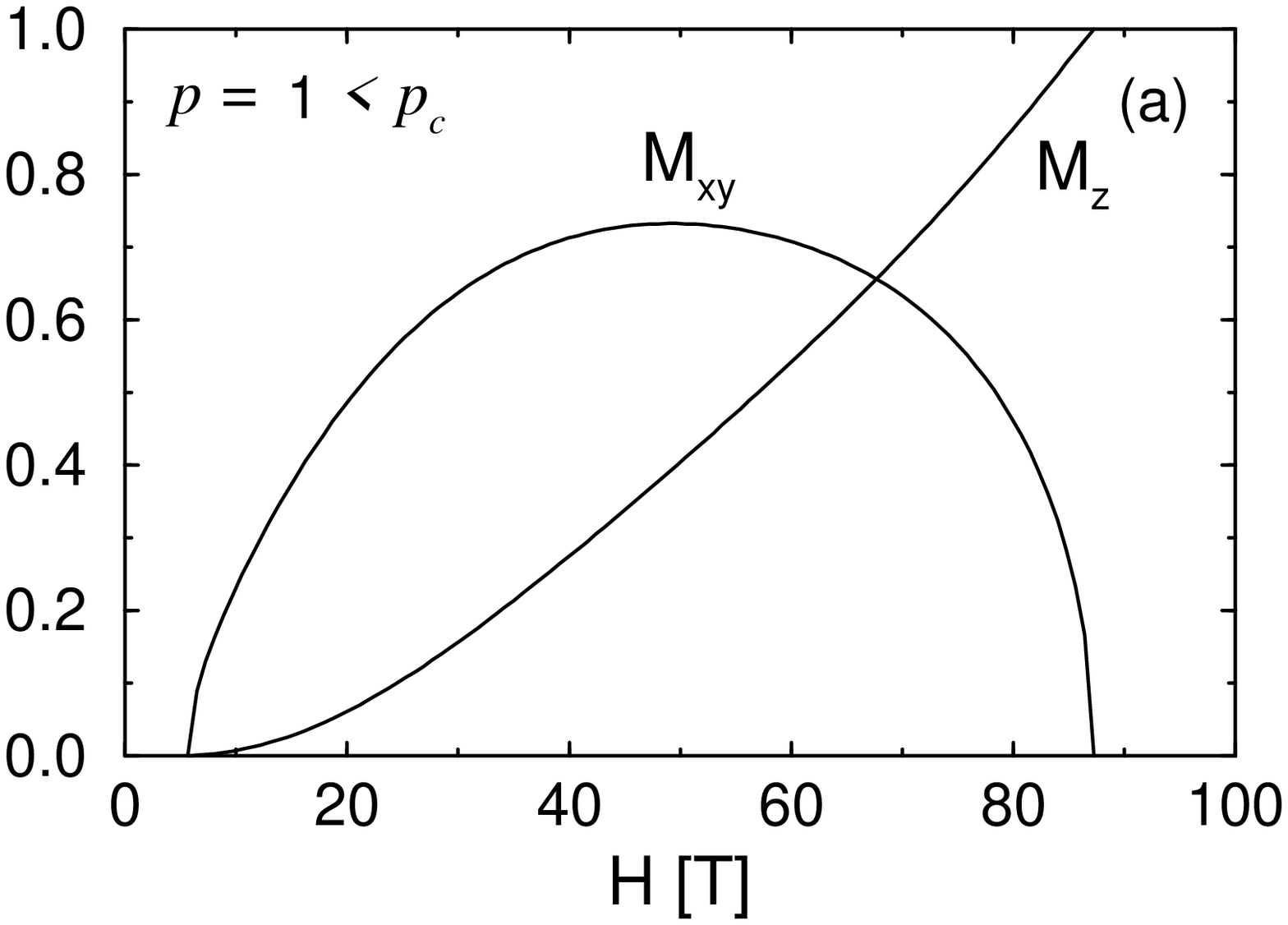}
\includegraphics[width=7cm]{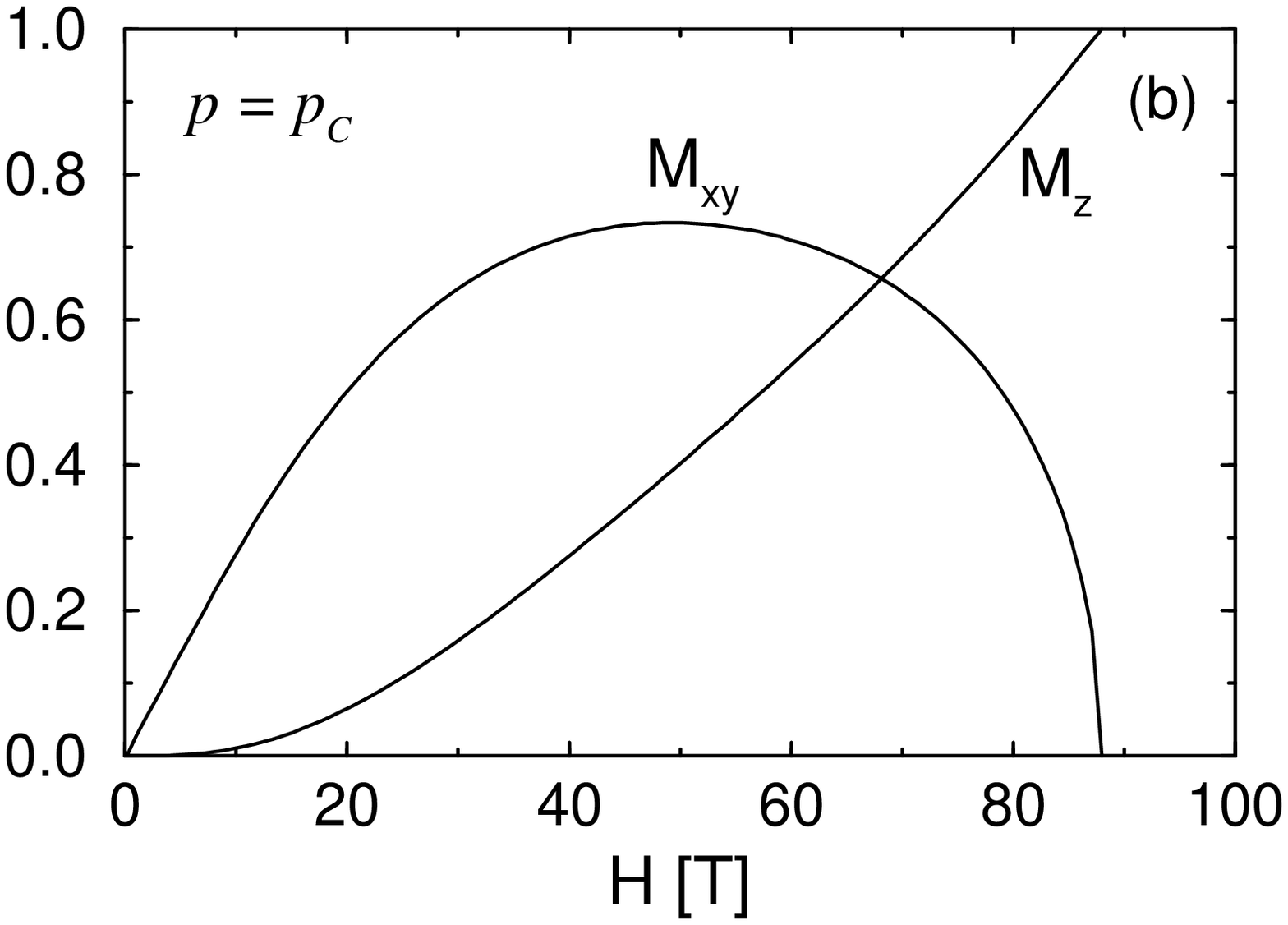}
\includegraphics[width=7cm]{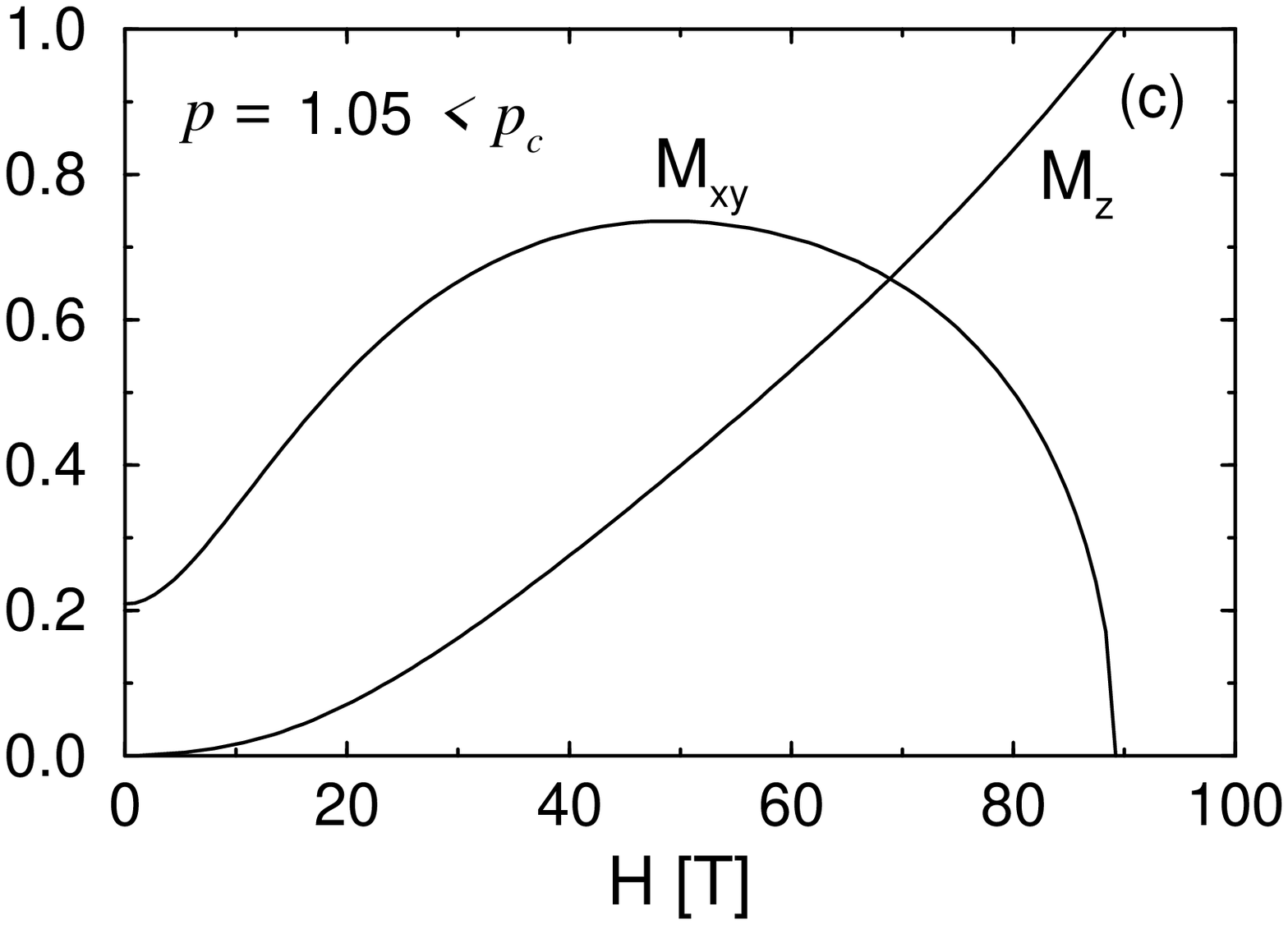}
\end{center}
\vspace{-0.8cm}
\caption{ Normalized magnetization curves for \Tl: (a) in the disordered phase
($p = 1 < p_c$); (b) at the critical pressure ($p = p_c$); (c) in the 
pressure--induced AF--ordered phase ($p = 1.05 > p_c$). $M_z$ and $M_{xy}$ 
denote respectively the uniform and staggered magnetizations, and we have 
taken a $g$--factor of 2.16.}
\label{fig:M-pressure-h}
\end{figure}

We conclude this section by discussing the uniform and staggered  
magnetization of \Tl~under finite pressure, for which we define the  
magnetization exponent near the critical field by  
\begin{equation}
\delta M = M(H) - M(H_c) \propto (H - H_c)^\gamma.
\end{equation}
Fig.~\ref{fig:M-pressure-h} shows the behavior of both magnetization
components for the three regimes $p < p_c$, $p = p_c$, and $p > p_c$.
In the disordered phase [Fig.~\ref{fig:M-pressure-h}(a)], the critical
field $h_c(p)$ is finite [Fig.~\ref{fig:phase-diagram}] and the uniform 
magnetization $M_z$ shows an apparent nonlinear field dependence over a 
significant field range, as discussed in Sec.~IIIB. However, close to the 
critical field the leading power--law dependence is linear $(\gamma = 1)$, 
reflecting the fact that the difference between \Tl~and \K~arising from 
stronger interdimer coupling is only quantitative in nature. The 
staggered moment, $M_{xy}$, has the mean--field exponent $\gamma = 1/2$
expected for a gapped phase (Sec.~IIIB). These exponents may also be
deduced by considering the dependence of the magnetizations on the
coefficient $v^2$ [Eqs.~(\ref{emfml}) and (\ref{emfmt})], which 
specifies the mixing of the triplet magnons into the singlet ground state 
(\ref{eqn:b}), given that $v^2 \propto H - H_c$ in the 
disordered phase. Increasing the pressure causes a decrease in $H_c(p)$ 
until it reaches zero at the critical pressure [$H_c(p_c) = 0$] as in 
Fig.~\ref{fig:phase-diagram}. In this case [Fig.~\ref{fig:M-pressure-h}(b)] 
both magnetizations develop from $H = 0$: the field induces a linear mixing 
of singlets and triplets in the ground state, $v \propto H$, and $M_z$ 
depends also on the difference $f^2 - g^2$ of the weights of the $t_+$ and 
$t_-$ modes [Eq. (\ref{emfml})], whose field--induced splitting is linear 
whence $\gamma = 3$. $M_{xy}$ is not sensitive to the latter splitting 
[Eq. (\ref{emfmt})], and depends only linearly on the singlet--triplet 
admixture, yielding $\gamma = 1$. Finally, the ordered phase at $p > p_c$ 
[Fig.~\ref{fig:M-pressure-h}(c)] has a finite staggered magnetization at 
zero field, and $v$ has a finite value at $H = 0$, leading to $\gamma = 1$ 
for $M_{xy}$. The uniform component appears to show a similar 
deviation from linearity as in the disordered phase, but for similar 
reasons the leading field dependence is in fact linear ($\gamma = 1$) 
close to $H = 0$.

The behavior of the two magnetization components in the vicinity of the 
critical fields is summarized by 
\begin{equation}
\delta M_z \simeq
\left\{
  \begin{array}{@{\,}lr}
    \frac{J - 2J'}{J'(3J - 4J')}(h - h_c) & ~~~~~~~~~~ (p < p_c) \\
    \frac{1}{(2J')^3}h^3 & ~~~(p = p_c) \\
    \frac{2J' - J}{(2J')^2}h & ~~~(p > p_c) \\
  \end{array}
\right\}
\label{eum}
\end{equation}
for the uniform component, and
\begin{equation}
\delta M_{xy} \simeq
\left\{
  \begin{array}{@{\,}lr}
    \left[ \frac{\sqrt{J^2 - 2JJ'}}{J'(3J - 4J')} \right]^{1/2} \!\!
    (h - h_c)^{1/2} & ~(p < p_c) \\
    \frac{1}{\sqrt{2}J'}h & ~(p = p_c) \\
    \frac{3J - 2J'}{8J'^2\sqrt{(2J')^2 - J^2}}
    h^2 & ~(p > p_c) \\
  \end{array}
\right\}
\label{esm}
\end{equation}
for the staggered component, where $J^\prime$ denotes the quantity
$|\epsilon_\bsQ|$ defined in Eq.~(\ref{ehqpopeq}). The magnetization
exponents are collected in Table~II.

\begin{table}[t!]
\begin{tabular}{|l|c|c|c|}
\hline
\multicolumn{1}{|r|}{} &  $\;p < p_c\;$   &  $\;p = p_c\;$  &  $\;p > p_c\;$ 
\\ \hline
 $M_z$ & $\; \gamma = 1 \;$ & $\; \gamma = 3 \;$ & $\; \gamma = 1 \;$  \\
 $M_{xy}$ & $\; \gamma = {\textstyle \frac{1}{2}} \;$ & $\; \gamma = 1 \;$ & 
$\; \gamma = 2 \;$  \\ \hline
\end{tabular}
\medskip
\caption{Exponent for magnetization curves.}
\end{table}

\section{Summary}

We have used a bond--operator model for TlCuCl$_3$ at zero temperature to  
provide a complete description of quantum phase transitions (QPTs) in  
interacting spin--dimer systems driven by both magnetic field and hydrostatic 
pressure. The magnetically ordered and disordered phases were characterized 
by analyzing both the dispersion relations of the triplet magnon modes and  
the staggered and uniform magnetizations as functions of field and pressure.  
At zero field and ambient pressure, a minimal coupled--dimer model gives a  
very good account of the magnon dispersions both in TlCuCl$_3$, despite 3d  
interdimer interactions which are almost sufficiently strong to close the  
spin gap, and in the structural analog KCuCl$_3$ where interdimer coupling  
is weaker. At finite fields the bond-operator framework provides a continuous  
description of the low--field, dimer--singlet phase, the intermediate--field  
regime of partial (staggered and uniform) magnetic order, and the fully  
polarized, classical phase at high fields. Under pressure the spin gap of  
the quantum dimer phase decreases monotonically to a single transition  
into a phase of staggered magnetic order. The evolution of the order  
parameters through the QCPs separating these regimes is continuous.  
 
Considering first the gapped system at ambient pressure, the magnetic  
excitations at low fields are merely the Zeeman--split magnons of the  
zero--field case, whose dispersion is unaltered by any field up to  
the lower critical field $H_{c}$, where the spin gap is closed. In the  
high--field regime above $H_s$, the excitations are precisely the gapped  
spin waves expected in a fully polarized antiferromagnet. At $H_{c}$,
a QPT takes place to an intermediate--field phase characterized by an  
antiferromagnetic order of the transverse component of the magnetization,  
and by one excitation mode which is massless at the Bragg point. This  
magnon has a ready interpretation as the phase fluctuation mode (Goldstone  
mode) of a condensate of mixed singlet and triplet character on each dimer  
rung for fields $H_{c} < H < H_{s}$. Because of its gapless nature the phase  
mode dominates the response of the system at intermediate  
magnetic fields, including most notably the field--dependence of the  
magnetization close to the QCPs. The bond--operator description of the  
ground state is a linear combination of singlet and triplets which  
changes continuously from pure singlet character at $H_c$ to a pure  
triplet state at $H_s$. The description of the phase mode is in good  
quantitative agreement with recent INS measurements performed at $H >  
H_{c}$.\cite{rrcfgkmwhv} 
 
If the QPT is interpreted as a destabilization of the disordered phase  
when one or more of its magnon modes become soft, a second route to an  
ordered phase is provided by the application of hydrostatic pressure to  
enhance the interdimer interactions and broaden the magnon band. In this  
situation the gapped magnons of the quantum dimer phase remain triply  
degenerate as the gap decreases, until at the quantum critical point one  
obtains a phase with three spin waves. A further increase of the interdimer  
couplings leads to a conventional antiferromagnet with two spin waves --  
gapless (Goldstone), transverse, phase modes of the staggered order  
parameter -- accompanied by one longitudinal mode which develops a gap  
and corresponds to amplitude fluctuations of the staggered moment. The  
appearance of pressure--induced order in \Tl~has recently been confirmed  
by experiments\cite{rofokt} which enable one to conclude that the critical  
pressure for the QPT is very small (a result now confirmed 
experimentally\cite{rtpc}), thus presenting a valuable additional  
control parameter for the experimental characterization of QPTs in the  
\Tl~system.  
 
Both magnetic field-- and pressure--induced ordering transitions can be
interpreted as a condensation of low--lying magnon modes, whose linear
combination with the singlet describes the ground state of the ordered
phase. Indeed, the fact that the modes become soft at wave vector $\bQ =
(0,0,2\pi)$ in both cases suggests a strong similarity between the QPTs
driven by both magnetic field and pressure. However, we have shown also 
that an important, if quantitative, difference exists concerning the 
degeneracy of the magnon modes and the breaking of rotational symmetry 
in the ordered phase. While the pressure--induced transition is characterized 
by two Goldstone modes and a spontaneous breaking of the full $O(3)$ symmetry 
of spin space, the field--induced transition has only one Goldstone mode 
corresponding to a spontaneous breaking of $O(2)$ symmetry in the plane 
normal to the applied field.

Returning to the description of the ordered magnetic phase as a Bose--Einstein
condensation (BEC) of magnons, the bond--operator formulation provides 
considerable additional insight into this approach. The Goldstone mode 
of the ordered phase corresponds to the linearly dispersive ``sound'' mode 
connected with BEC. The associated creation and annihilation processes arise 
from the triplet--triplet interaction ${\cal H}_{\rm tt}$, which in the 
ordered phase generates terms of the form ${\overline a}^2 b_{\bsk +} 
b_{-\bsk +}$ [Eq.~(\ref{eob2hpm})] as a consequence of the condensation of 
the mode $t_+$. A recent numerical analysis of the scaling exponents of the 
magnetization in the ordered regime\cite{rnwnh} has confirmed that the BEC 
scenario provides a suitable description of the QPT, and is universally 
valid for all interdimer couplings. We emphasize, however, that away from 
the transition it is necessary to modify the ground--state condensate, to a 
linear combination of the dimer singlet and two of the triplet components,
in order to describe the ordered phase across the entire parameter region.

The theoretical analysis of QPTs is often limited by the absence of  
suitable descriptions valid simultaneously for the phases on both sides  
of the transition. Approximate approaches applied in one phase or the  
other are fated not to converge on a single transition point, and thus  
cannot provide a continuous evolution in physical properties across the  
transition. The bond--operator formulation for the magnetic QPTs in  
spin--dimer systems is appropriate for both disordered and ordered  
phases, and thus represents an exceptional case illustrating the continuity  
of physical properties across such transitions. Here we note, however, in 
the context of \Tl, that two recent, high--resolution experimental 
studies\cite{rslbot,rvtvot} have revealed a small discontinuity at the 
ordering transition. The weakly first--order magnetic phase transition 
is accompanied by a simultaneous structural distortion, presumably as a 
consequence of magnetoelastic coupling which is not included in the present 
analysis. The nature of this coupled transition awaits further investigation. 
We comment briefly that the remarkable success of the bond--operator framework 
for the global phase diagrams of \Tl~and \K~rests to a considerable extent 
on the fact that the dimer system has significant coupling in all three 
spatial dimensions, as a result of which the mean--field approximations 
employed here are highly accurate. This is reflected in the large value 
of the singlet condensate parameters in the disordered phases of \Tl~and \K, 
which deviate from unity due to quantum fluctuations by at most 3\%. The 
dimensionality of the system is also responsible for the success of the 
approximation that the constraint is applied only globally. This approach 
remains appropriate if it is possible for dimer triplets, which are 
effectively hard--core bosons, to avoid each other by perpendicular hopping 
processes. Only in one dimension (1d) is this no longer possible, and so the 
bond--operator treatment of the isolated spin ladder would require a more 
systematic application of the local constraint in order to reproduce the 
incommensurate excitations unique to 1d.
 
We close by summarizing the experimental possibilities offered by the  
study of QPTs in magnetic systems. While some experiments are now under  
way to investigate the field--induced QPT, most notably by inelastic  
neutron scattering to measure the magnon modes both below and above the  
transition, much scope remains for deeper analysis by techniques  
including NMR. The application of pressure offers a very promising field  
for future studies. The experimental result of robust magnetic order at  
high pressure shows that it is possible to achieve a genuine tuning of  
superexchange constants through the QPT by the application of hydrostatic  
pressure. Further, the rather small pressures required imply the  
possibility of a variety measurements, including pressure--dependent  
specific heat, magnetization, magnetic susceptibility, elastic neutron  
scattering for the magnetic order parameter and inelastic neutron  
scattering for magnon dispersion relations. All of these quantities  
may be measured through the continuous QPT, and in combination with  
ongoing field--dependent studies yield the possibility of obtaining a  
complete experimental map of the QPT from magnetic order to quantum  
disorder in \Tl. We have provided a number of predictions for the  
properties of the phases in this map, including most notably the  
evolution of the magnon dispersion relations and the magnetization  
exponents furnished in Table~II. In this connection we also stress the  
possibility of performing experiments to detect the amplitude mode in  
the pressure--induced ordered phase, which should be visible in  
susceptibility and NMR measurements, as well as by inelastic neutron  
scattering. In conclusion, the phenomena of field-- and pressure--induced 
magnetic order provide excellent examples of quantum phase transitions 
readily controlled by experiment.  
 
\acknowledgments 
 
We are indebted to N. Cavadini, H. Kusunose, F. Mila, A. Oosawa, 
Ch. R\"{u}egg, and H. Tanaka for many valuable discussions. We thank 
N. Cavadini and Ch. R\"{u}egg for the helpful provision of Fig.~1. This 
work was supported by the Japan Society for the Promotion of Science 
(JSPS) and the MaNEP project of the Swiss National Science Foundation. 
 
\appendix* 
 
\section{} 
 
Here we present the expressions for $\epsilon_{\bsk i}$ and $\Delta_{\bsk i}$ 
used in the transformed Hamiltonian (\ref{eqn:hamiltonian-order}) at  
quadratic order in the triplet modes (\ref{eob2hpm}). The energies are  
given by  
\begin{eqnarray} 
\epsilon_{\bsk +} & = & (\epsilon_+ f^2 +\epsilon_- g^2)(u^2 - v^2) + C_0 
- 2 u^2 v^2 \epsilon_\bsQ \label{eaope} \nonumber \\ 
& & + (u^4+v^4)\epsilon_\bsk - u^2 v^2 [4 f g + (f^2 - g^2)^2] 
(\epsilon_\bsk+\epsilon_\bsQ), \nonumber \\ 
\epsilon_{\bsk 0} & = & \epsilon_0 - v^2(\epsilon_+ f^2 + \epsilon_- g^2) 
+ C_0                         + (u^2 - v^2) \epsilon_\bsk, \nonumber \\ 
\epsilon_{\bsk -} & = & \epsilon_+ g^2 + \epsilon_- f^2 - v^2(\epsilon_+  
f^2 + \epsilon_3 g^2) \\ 
& & + v^2 (f^2 - g^2)^2 \epsilon_\bsQ + C_0 + (u^2 - 4 f^2 g^2 v^2)  
\epsilon_\bsk, \nonumber \\ 
\epsilon_{\bsk \pm} & = & u( -\epsilon_+ + \epsilon_- ) f g \nonumber \\ 
& & - u v^2 (1 - 2 f g)(f^2 - g^2)(\epsilon_\bsk + \epsilon_\bsQ), \nonumber \\
\epsilon_\alpha &=& J - \alpha h,~~~(\alpha=+,0,-) \nonumber
\end{eqnarray} 
where $\epsilon_\bsk$ is given by Eq.~(\ref{eepsk2}) and  
\begin{equation} 
C_0 = - \epsilon_\bsQ [2 u^2 v^2 (1 + 2fg) - v^4 (f^2 - g^2)^2], 
\end{equation} 
and the coefficients of the pairing terms by  
\begin{eqnarray} 
\Delta_{\bsk +} & = & -[2 u^2 v^2 - 2 f g (u^4 + v^4) + u^2 v^2 (f^2 -  
g^2)^2] \epsilon_\bsk, \label{eaopd} \nonumber \\ 
\Delta_{\bsk 0} & = & (u^2 + 2 f g v^2)\epsilon_\bsk, \nonumber \\ 
\Delta_{\bsk -} & = & - (u^2 + v^2 2 f g) 2 f g \epsilon_\bsk, \\ 
\Delta_{\bsk \pm} & = & (u^3 + 2 f g u v^2 )(f^2-g^2)\epsilon_\bsk.\nonumber  
\end{eqnarray}

\end{document}